\documentclass[12pt,draftcls,onecolumn]{IEEEtran}

\usepackage{amsfonts}
\usepackage{cite}
\usepackage{graphicx}
\usepackage{amssymb}
\usepackage{subfigure}
\usepackage{array}
\usepackage{color}
\usepackage{amsmath}
\usepackage{algorithmic}
\usepackage{algorithm}
\usepackage{cases}
\usepackage{mathrsfs}
\usepackage{epsfig}
\usepackage{psfrag}
\usepackage{url}
\usepackage{setspace}
\usepackage{stfloats}
\usepackage{enumerate}
\allowdisplaybreaks[4]

\usepackage{etoolbox}
\makeatletter
\patchcmd{\@makecaption}
  {\scshape}
  {}
  {}
  {}
\makeatletter
\patchcmd{\@makecaption}
  {\\}
  {.\ }
  {}
  {}
\makeatother

\newcommand{\dis}{\stackrel{d}{\sim}}

\newtheorem{Thm}{Theorem}
\newtheorem{Lem}{Lemma}

\newtheorem{Prob}{Problem}

\IEEEoverridecommandlockouts

\newcommand{\blue}[1]{\textcolor{black}{#1}}

\begin{document}

\title{Optimal Caching Designs for Perfect, Imperfect and Unknown File Popularity Distributions in Large-Scale Multi-Tier Wireless Networks}

\author{\IEEEauthorblockN{Chencheng Ye, Ying Cui, Yang Yang and Rui Wang\thanks{C. Ye and Y. Cui are with Shanghai Jiao Tong University, China. Y. Yang is with University of Luxembourg, Luxembourg. R. Wang is with Southern University of Science and Technology, China.}}}




\maketitle


\vspace{-11mm}
\begin{abstract}
Most existing caching solutions for wireless networks rest on \blue{the ideal} assumption that the file popularity distribution is perfectly known. In this paper, we consider optimal \blue{random} caching designs for perfect, imperfect and unknown file popularity distributions in \blue{a} large-scale multi-tier wireless network.
First, in the case of perfect file popularity distribution, we formulate \blue{the successful transmission probability (STP) optimization problem}, which is nonconvex. We \blue{propose} an efficient parallel iterative algorithm to obtain a stationary point \blue{based on} parallel successive convex approximation (SCA).
Then, in the case of imperfect file popularity distribution, we formulate \blue{the worst-case STP maximization problem.} To solve this challenging robust optimization problem, we transform it to an equivalent complementary geometric programming (CGP), and \blue{propose} an efficient iterative algorithm \blue{to obtain} a stationary point \blue{based on} SCA.
To the best of our knowledge, this is the first work explicitly considering the estimation error of file popularity distribution in the optimization of caching design. 
Next, in the case of unknown file popularity distribution, we formulate \blue{the stochastic STP (i.e., the STP in the stochastic form) maximization problem. This} is a challenging nonconvex stochastic optimization problem, \blue{and} we develop an efficient iterative algorithm to obtain a stationary point \blue{based on} stochastic parallel SCA. As far as we know, this is the first work considering stochastic optimization in a large-scale wireless network. 
Finally, by numerical results, we show that the proposed solutions achieve \blue{notable} gains over existing schemes in all three cases, \blue{and reveal the values of the robust caching optimization and stochastic caching optimization in the cases of imperfect file popularity distribution and unknown file popularity distribution, respectively.} 
\end{abstract}


\section{Introduction}

The rapid proliferation of smart mobile devices has triggered an unprecedented growth of the global mobile data traffic. \blue{Caching content closer to end users, e.g., at base stations (BSs) \cite{cooperative,wanli,TassiulasIT13,ICC15Giovanidis,cui2016analysis,femtocaching13,cui2017analysis,li2016optimization,wen2016cache,Wang2017Joint} or even at end users~\cite{centralali,maddah2015decentralized}}, has been proposed as an effective way to support the dramatic traffic growth, by reducing the distance between popular contents and requesters, and alleviating the backhaul load.
\blue{In addition, caching has also been jointly designed with multicast~\cite{cui2016analysis,cui2017analysis,Wang2017Joint} and cooperation \cite{cooperative,wanli}. In this paper, we focus on caching at BSs, and a large body of recent research would be out of the scope of this paper.}

Caching in single-tier wireless networks has been actively studied~\cite{TassiulasIT13,ICC15Giovanidis,cui2016analysis}. 
Specifically, in~\cite{TassiulasIT13}, the authors consider the optimal caching design and transmission strategy to minimize the required link capacity in a square grid wireless network.
In~\cite{ICC15Giovanidis} and~\cite{cui2016analysis}, the authors consider random caching at BSs, analyze and optimize the hit probability~\cite{ICC15Giovanidis} and the successful transmission probability (STP)~\cite{cui2016analysis} in large-scale wireless networks that capture the stochastic nature of geographic locations of BSs and users.\footnote{
\blue{The stochastic network model is a widely used tractable model that allows to study the average behavior of a network using mathematical tools from stochastic geometry and is about as accurate as the standard grid model, when compared to an actual network~\cite{Andrews11}.}}
As caching can successfully alleviate the urgent backhaul requirement for small cells, a significant amount of research effort has been devoted to optimal caching design in multi-tier wireless networks~\cite{femtocaching13,cui2017analysis,li2016optimization,wen2016cache,Wang2017Joint}. For instance, in~\cite{femtocaching13}, the authors consider coded and uncoded caching at small BSs to minimize the expected downloading time in a macro cell with multiple small BSs \blue{and users at fixed locations}.
In~\cite{cui2017analysis,li2016optimization,wen2016cache,Wang2017Joint}, the authors consider hybrid caching~\cite{cui2017analysis} and random caching~\cite{li2016optimization,wen2016cache,Wang2017Joint} in large-scale multi-tier wireless networks, and focus on the analysis and optimization of the STP. Specifically, in our previous work~\cite{cui2017analysis}, we obtain optimal hybrid caching design for a two-tier wireless network. 
In~\cite{li2016optimization} and~\cite{wen2016cache}, the authors obtain optimal caching design for a multi-tier network in the case of uniform signal-to-interference ratio (SIR) threshold for all users. In the general case of arbitrary SIR thresholds for users, the optimization problem is nonconvex, and in~\cite{wen2016cache}, an optimal caching solution of a simplified convex problem is used as a sub-optimal solution of the original nonconvex problem. In our previous work~\cite{Wang2017Joint}, a stationary point of the nonconvex problem is obtained only for a two-tier wireless network.
Note that most existing works on caching~\cite {cooperative,wanli,TassiulasIT13,ICC15Giovanidis,cui2016analysis,femtocaching13,cui2017analysis,li2016optimization,wen2016cache,Wang2017Joint,centralali,maddah2015decentralized} assume that the file popularity distribution is perfectly known. In practice, however, such an assumption cannot be reasonably justified~\cite{Tatar2014}.

Some recent works consider caching design in the case \blue{that} the file popularity distribution is not known and only instantaneous file requests from users can be observed\cite{bacstuug2015transfer,Leconte,Bharath,Chenyang,Trend,cachingmimoLiu15,blasco2014learning,song,reinforcement,JSTSP18}. These works generally fall into two categories. One category adopts two-stage methods for caching design in the case of unknown file popularity distribution~\cite{bacstuug2015transfer,Bharath,Chenyang,Trend,Leconte}. 
Specifically, in the first stage, the file popularity distribution is estimated based on historical file requests, via various learning approaches; in the second stage, caching schemes are proposed based on the estimated popularity distribution. 
To be specific, in~\cite{bacstuug2015transfer} \blue{and~\cite{Trend}}, the file popularity distribution is estimated using transfer learning~\cite{bacstuug2015transfer} \blue{and file feature space partitioning~\cite{Trend}.} Based on the estimated file popularity distribution, the authors consider caching the most popular files, and analyze the backhaul offloading~\cite{bacstuug2015transfer} \blue{and cache hit~\cite{Trend}.}
In~\cite{Bharath}, the request frequency for each file obtained from historical file requests is considered as the popularity of the file, and the performance gap for minimizing the offloading time caused by the estimation error is analyzed.
In~\cite{Chenyang}, the file popularity distribution is estimated by learning user preferences via probabilistic latent semantic analysis, and a greedy algorithm is proposed to obtain a caching solution of the offloading probability maximization problem (under the estimated file popularity distribution) with performance guarantee.
\blue{In~\cite{Leconte}, given apriori information on the file popularity distribution, the file request rate is estimated using Bayesian framework, and an asymptotically optimal policy is proposed to maximize the hit probability (under the estimated file popularity distribution).}
Note that these works~\cite{bacstuug2015transfer,Bharath,Chenyang,Trend,Leconte} consider either simple caching design without performance guarantee, such as caching the most popular files at each BS~\cite{bacstuug2015transfer,Trend}, or optimization-based caching design obtained by optimizing simple performance metrics that may not fully reflect natures of wireless networks (such as fading~\cite{Bharath} and stochastic locations of BSs and users~\cite{Chenyang,Leconte}). In addition, note that~\cite{bacstuug2015transfer,Bharath,Chenyang,Trend,Leconte} fail to consider estimation errors in designing caching schemes.

The other category adopts single-stage methods for caching design~\cite{cachingmimoLiu15,blasco2014learning,song,reinforcement,JSTSP18} in the case of unknown file popularity, where caching solutions are gradually updated while accumulating file requests using stochastic optimization~\cite{cachingmimoLiu15} or reinforcement learning~\cite{blasco2014learning,song,reinforcement,JSTSP18}. 
Compared with two-stage methods, \blue{the caching solutions obtained by single stage methods can be continuously improved as new file requests are observed.}
Specifically, in~\cite{cachingmimoLiu15}, the authors optimize power control and caching for video streaming in a multi-cell multi-user MIMO network using techniques for stochastic optimization.
In~\cite{blasco2014learning} and~\cite{song}, the authors formulate the optimal caching design problem for a single-cell wireless network~\cite{blasco2014learning} and the optimal cooperative caching design problem for a multi-cell wireless network~\cite{song}, and develop low-complexity algorithms to obtain approximate solutions using results for multi-armed bandit problems.
In~\cite{reinforcement} and~\cite{JSTSP18}, the authors formulate the dynamic optimal caching design problem for a single-cell wireless network~\cite{reinforcement} and a multi-cell wireless network~\cite{JSTSP18}, and develop approximate solutions using Q-learning techniques.
\blue{However, the fixed network topologies considered in~\cite{cachingmimoLiu15,blasco2014learning,song,JSTSP18,reinforcement} cannot fully capture the geographic features of the locationsof BS and users.}
In addition, the caching solutions in~\cite{cachingmimoLiu15,blasco2014learning,song,JSTSP18,reinforcement} are for one BS~\cite{blasco2014learning,reinforcement} or a single-tier of BSs~\cite{cachingmimoLiu15,song,JSTSP18}.

Therefore, further studies are \blue{needed} to optimize caching design in multi-tier wireless networks when perfect file popularity distribution is not known. In this paper, we consider optimal \blue{random} caching design in three cases, i.e., the case of perfect file popularity distribution (where the file popularity distribution has been estimated, and the estimation error is negligible), the case of imperfect file popularity distribution (where the file popularity distribution has been estimated, and a deterministic bound of the estimation error is known) and the case of unknown file popularity distribution (where there is no prior information of the file popularity distribution, but instantaneous file requests from some users can be observed over time), in a large-scale multi-tier wireless network.
Our main contributions are summarized below.
\begin{itemize}
\item In the case of perfect file popularity distribution, we formulate the STP maximization problem, which is nonconvex with a complicated objective function. We \blue{propose} an efficient parallel iterative algorithm to obtain a stationary point \blue{based on} parallel successive convex approximation (SCA)~\cite{razaviyayn2014parallel}. Specifically, by carefully \blue{designing} an approximation function for each tier, we obtain closed-form optimal solutions \blue{for} the approximate \blue{functions of} all tiers at each iteration, and hence can significantly reduce the \blue{computational} complexity and improve the convergence speed of the iterative algorithm.

\item In the case of imperfect file popularity distribution, we formulate the worst-case STP (over all possible values of the true file popularity distribution) maximization problem. \blue{This} is a challenging \blue{robust optimization} problem which does not lie in the category of convex-concave games that can be easily solved. We transform it to an equivalent complementary geometric programming (CGP) and \blue{propose} an efficient iterative algorithm \blue{to obtain} a stationary point using SCA~\cite{4275017}.
Note that this case corresponds to that considered in the second stage of the two-stage methods proposed in \cite{bacstuug2015transfer,Bharath,Chenyang,Trend,Leconte}. But the essential difference is that we explicitly consider the estimation error of file popularity distribution in the optimization of caching design.
\item In the case of unknown file popularity distribution, we formulate the stochastic STP (i.e., the STP in the stochastic form) maximization problem. \blue{This} is a challenging nonconvex stochastic optimization problem, \blue{and} we develop an efficient parallel iterative algorithm to obtain a stationary point \blue{based on} stochastic parallel SCA~\cite{7412752}. Specifically, by carefully \blue{designing} an approximation function for each tier, we obtain closed-form optimal solutions \blue{for} the approximate \blue{functions of} all tiers at each iteration, make full use of instantaneous file requests from users at each slot, and significantly improve the convergence speed of the iterative algorithm. 
Note that this case corresponds to that considered in the single-stage methods proposed in \cite{cachingmimoLiu15,blasco2014learning,song,reinforcement,JSTSP18}. The key difference is that we consider stochastic optimization in a large-scale wireless network which captures the channel fading, interference and stochastic nature of wireless networks.
\item Finally, by numerical simulations, we show the convergence of the proposed solutions. We also show that the proposed solutions achieve \blue{notable} gains over existing schemes in all three cases. 
\end{itemize}

\section{System Model}\label{sec:netmodel}

\subsection{Network Model}

\begin{figure}[t]
\begin{center}
\includegraphics[width=12cm]{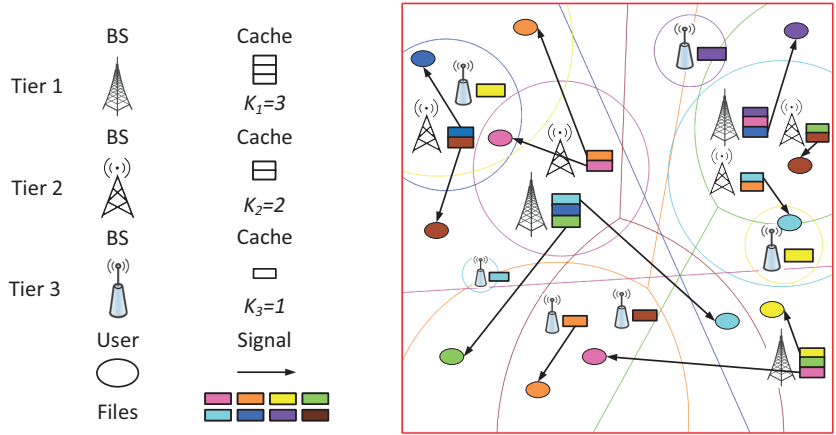}
\end{center}
\vspace{-3mm}
\caption{\small{Network model. $M=3$ and $N=8$. Each file $n\in \mathcal N$ corresponds to a Voronoi tessellation (in the same color as the file),  determined  by the locations and transmission powers of all BSs storing this file.}}
\label{fig:system}
\vspace{-3mm}
\end{figure}

In this part, we first elaborate on the network model which extends those in~\cite{li2016optimization,wen2016cache,Wang2017Joint} in the sense that besides perfect file popularity distribution, it also models imperfect and unknown file popularity distributions. 
We consider a large-scale $M$-tier network consisting of $M$ tiers of BSs, where $M\geq2$,\footnote{\blue{Note that for $M=1$, the case of perfect file popularity distribution has been studied in our previous work~\cite{cui2016analysis}, and the cases of imperfect and unknown file popularity distributions can be easily investigated following the results for $M\geq2$ in this paper.}}
as shown in Fig.~\ref{fig:system}.
The locations of the BSs in tier $m$ are spatially distributed as an independent homogeneous Poisson point process (PPP), denoted as $\Phi_{m}$, with density $\lambda_{m}$, for all $m\in\mathcal{M}\triangleq\{1,2,\dots,M\}$. The locations of the users are also distributed as an independent homogeneous PPP $\Phi_{u}$. 
Consider a discrete-time system with time being slotted. Let $t\in\{1,2,\dots\}$ denote the slot index.
Each BS in tier $m$ has one transmit antenna with \blue{fixed} transmission power $P_m$. Each user has one receive antenna. All BSs are operating on the same frequency band \blue{and each BS adopts an orthogonal transmission mechanism over frequency or time at each slot.} Both path loss and small-scale fading are considered: for path loss, a transmitted signal from either tier with distance $D$ is attenuated by a factor $D^{-\alpha}$, where $\alpha>2$ is the path loss exponent~\cite{cui2017analysis,li2016optimization,wen2016cache,Wang2017Joint}; for small-scale fading, at each slot, Rayleigh fading channels are adopted. Since a multi-tier network is primarily interference-limited, we ignore the thermal noise for simplicity\cite{wen2016cache}.

Let $\mathcal N\triangleq \{1,2,\cdots, N\}$ denote the set of $N$ files in the $M$-tier network.
For ease of illustration, as in \cite{femtocaching13,cui2017analysis,li2016optimization,wen2016cache,Wang2017Joint,bacstuug2015transfer},  assume that all  files  have the same size.\footnote{The results in this paper can be easily extended to the case of different file sizes by considering file combinations of the same total size, but formed by files of possibly different sizes \cite{Wang2017Joint}.}
At each slot, a user requests at most one file at random. \blue{Let $\pi_s(t)\in\{0\}\cup\mathcal{N}$ denote the file request status of user $s\in\Phi_u$ at slot $t\in\{1,2,\dots\}$, where $\pi_s(t)=0$ if user $s$ does not request any file and $\pi_s(t)=n\in\mathcal N$ if user $s$ requests file $n$.}
For ease of analysis, assume $\pi_s(t)$, $s\in\Phi_u$, $t\in\{1,2,\dots\}$ are i.i.d. with respect to $s$ and $t$~\cite{cui2017analysis,li2016optimization,wen2016cache,Wang2017Joint,bacstuug2015transfer}.\footnote{
\blue{The results in this paper can be extended to the case where there are multiple classes of users and file requests of users in the same class are i.i.d..}}
\blue{
Denoted $a_n\triangleq{\rm Pr}[\pi_s(t)=n|\pi_s(t)\neq0]\in [0,1]$, $s\in\Phi_u$, $n\in\mathcal N$, where $\sum_{n\in \mathcal N}a_n=1$.}
Thus, $\mathbf a\triangleq (a_n)_{n\in \mathcal N}$ represents the file popularity distribution, which usually evolves at a slower timescale. 
In this paper, we consider the following three cases of file popularity distribution. \\
\textbf{Perfect file popularity distribution:} In this case, we assume that the file popularity distribution has been estimated by some learning methods, and the estimation error is negligible. That is, the exact value of $\mathbf{a}$ is known.\\
\textbf{Imperfect file popularity distribution:} In this case, we assume that the file popularity distribution has been estimated by some learning methods, and a deterministic bound of the estimation error is known. Note that this case corresponds to that considered in the second stage of the two-stage methods proposed in \cite{bacstuug2015transfer,Bharath,Chenyang}. Let $\widehat{\mathbf a}\triangleq(\widehat a_n)_{n\in\mathcal N}$ denote the estimated file popularity distribution, and let $\boldsymbol\Delta\triangleq(\Delta_n)_{n\in\mathcal N}$ denote the estimation error. Assume $\sum_{n\in\mathcal N}\widehat a_n=1$, $\sum_{n\in\mathcal N}\Delta_n=0$, and $|\Delta_n|\leq\varepsilon_n$ for some known $\varepsilon_n>0$, for all $n\in\mathcal N$, which are usually satisfied for effective learning methods. The (true) file popularity distribution is given by $\mathbf a=\widehat{\mathbf a}+\boldsymbol\Delta$, and satisfies $\mathbf a\in\mathcal A \triangleq\left\{ (x_n)_{n\in\mathcal{N}} \ \middle|\ \underline a_n\leq x_n\leq \overline a_n, n\in\mathcal{N}, \sum_{n\in\mathcal{N}}x_n=1 \right\}$, where $\underline a_n\triangleq\max\{\widehat a_n-\varepsilon_n,0\}$, and $\overline a_n\triangleq\min\{\widehat a_n+\varepsilon_n,1\}$ for all $n\in\mathcal N$.\\
\textbf{Unknown file popularity distribution:} In this case, we assume that there is no prior information of the file popularity distribution $\mathbf{a}$, but the file requests from the users in some set $\mathcal U\subseteq\Phi_u$ \blue{without bias} can be observed over time. 
Here $\mathcal U$ can represent the set of users located \blue{in the physical area of} a BS or a cluster of BSs.\footnote{
\blue{For example, a user can submit its file request to its nearest BS, and each BS can gather file requests corresponding to an unbiased observation of the file popularity distribution.}
} Note that this case corresponds to that considered in the single-stage methods proposed in \cite{cachingmimoLiu15,blasco2014learning,song,reinforcement,JSTSP18}.

The $M$-tier network consists of cache-enabled BSs. In tier $m$, each BS is equipped with a cache of size $K_m<N$ (in number of files) to store $K_m$ different popular files out of $N$.
We say every $K_m$ different files form a combination. Thus, there are in total $\binom{N}{K_m}$ different combinations, each with $K_m$ different files. Let $\mathcal I_m$ denote the set of indices for the \blue{$\binom{N}{K_m}$} combinations. More detailed descriptions of $\mathcal{I}_m$ and $\mathcal{I}_{m,n}$ can be found in~\cite{cui2016analysis,cui2017analysis,Wang2017Joint}.



\subsection{Caching and User Association}
To provide high spatial file diversity, we consider random caching in the cache-enabled $M$-tier network~\cite{li2016optimization,wen2016cache,Wang2017Joint}, as illustrated  in Fig.~\ref{fig:system}. In particular, each BS in tier $m$ stores $K_m$ different files with certain probability~\cite{li2016optimization,wen2016cache,Wang2017Joint}.
The probability that combination $ i\in\mathcal I_m $ is stored \blue{at} each BS \blue{in} tier $m$ is $ p_{m,i} $,
where $p_{m,i}$ satisfies\blue{:}
\begin{align}
&0\leq p_{m,i}\leq1, \ m\in\mathcal{M}, \quad i\in \mathcal I_m,\label{eqn:p-interval}\\
&\sum_{i\in \mathcal I_m}p_{m,i}=1,\quad m\in\mathcal{M}.\label{eqn:p-sum1}
\end{align}
A random caching design is specified by the caching distributions \blue{for file combinations $\mathbf p\triangleq (p_{m,i})_{m\in\mathcal M, i\in \mathcal I_m}$.} Let $\mathcal I_{m,n}\subset\mathcal I_m$ denote the set of indices for the $\binom{N-1}{K_m-1}$ combinations containing file $n$. 
Based on $\mathbf{p}$, we also define the probability that file $n$ is stored at a BS in tier $m$, i.e., 
\begin{align}
T_{m,n} \triangleq \sum_{i\in\mathcal {I}_{m,n}}p_{m,i}, \quad m\in\mathcal{M},\ n \in \mathcal N. \label{eqn:T-def}
\end{align}
From \cite{ICC15Giovanidis,cui2016analysis,cui2017analysis,Wang2017Joint}, we know that the constraints on $ \mathbf p$ in \eqref{eqn:p-interval}, \eqref{eqn:p-sum1} and \eqref{eqn:T-def} can be equivalently  rewritten as the following constraints on $\mathbf T\triangleq(T_{m,n})_{m\in\mathcal M,n\in\mathcal N}$:
\begin{align}
&0\leq T_{m,n}\leq 1,\quad m\in\mathcal M,\ n\in\mathcal N, \label{eqn:T1}\\
&\sum\limits_{n\in\mathcal N}T_{m,n}=K_m,\quad m\in\mathcal M. \label{eqn:T2}
\end{align}
The constraints in~\eqref{eqn:T2} are due to the fact that each file combination in tier $m$ contains $K_m$ different files and the sum of the caching probabilities \blue{for} all file combinations is one. The details can be found in~\cite{cui2016analysis}. For any $\mathbf T$ satisfying \eqref{eqn:T1} and \eqref{eqn:T2}, one corresponding \blue{$\mathbf p$ satisfying \eqref{eqn:p-interval}, \eqref{eqn:p-sum1} and \eqref{eqn:T-def}} can be easily obtained using the method \blue{proposed} in \cite{ICC15Giovanidis}.\footnote{\blue{When the file popularity changes, the cache content of each BS can be easily updated based on its previous cached content.}}
\blue{Random caching design is specified by the caching distributions for file combinations $\mathbf p$, the size of which is $\sum_{m=1}^M\binom{N}{K_m}$. However, later we shall see that the performance metrics and optimal random caching design problems considered in this paper depend only on the caching probabilities for files $\mathbf T$, the size of which is $MN$.}

If a file is stored in a tier, a user requesting the file is associated with the BS which provides the maximum long-term average received power among all BSs storing the file~\cite{li2016optimization,wen2016cache,Wang2017Joint}. Otherwise, the user will be served through other service mechanisms~\cite{li2016optimization,wen2016cache,Wang2017Joint}, the investigation of which is beyond the scope of this paper.\footnote{
\blue{As in~\cite{li2016optimization,wen2016cache,Wang2017Joint}, we assume that user association can be done through some signaling mechanisms. For example, a user can submit its file request to its nearest BS and associates with its serving BS via the help of its nearest BS.}}
The probability that an arbitrary user requesting file $n$ is associated with tier $m$ is given \blue{by}~\cite{li2016optimization,wen2016cache,Wang2017Joint}:
\begin{align}
&A_{m,n}(\mathbf T) 
=\frac{\lambda_mT_{m,n}}{\lambda_m T_{m,n} + \sum_{l\in\mathcal{M}\backslash\{m\}}\lambda_{l}T_{l,n}\left(\frac{P_l}{P_m}\right)^{ \frac{2}{\alpha} }}, \quad
m\in\mathcal M,n\in\mathcal N. \label{eqn:user-association-prob}
\end{align}

\subsection{Performance Metrics}\label{Sec:perf}
In this paper, we study w.l.o.g. the performance of a typical user $u_0$, which is located at the origin. Suppose $u_0$ requests file $n$ and is associated with tier $m$. Let $\ell_0\in\Phi_{m}$ denote the index of the serving BS of $u_0$.
We denote $D_{m',\ell,0}$ and $h_{m',\ell,0}\dis \mathcal{CN}\left(0,1\right)$ as the distance and the small-scale channel between BS $\ell\in\Phi_{m'}$ and $u_{0}$, respectively.
For analytical tractability, as in \cite{wen2016cache,Wang2017Joint}, we assume all BSs are active for serving their own users.\footnote{This assumption corresponds to the worst-case interference strength for the typical user. The performance obtained under this assumption provides a lower bound on the performance of the practical network where some void BSs may be shut down.} 
In this case, the SIR of $u_{0}$, denoted by ${\rm SIR}_{m,n,0}$, is given by~\cite{Wang2017Joint}:
\begin{align}
&{\rm SIR}_{m,n,0}=\frac{{D_{m,\ell_0,0}^{-\alpha}}\left|h_{m,\ell_0,0}\right|^{2}}{\sum\limits_{\ell\in\Phi_{m}\backslash \{\ell_0\}}D_{m,\ell,0}^{-\alpha}\left|h_{m,\ell,0}\right|^{2}+\sum\limits_{j\in\mathcal{M}\backslash \{m\}}\sum\limits_{\ell\in\Phi_{j}}D_{j,\ell,0}^{-\alpha}\left|h_{j,\ell,0}\right|^{2}\frac{P_{j}}{P_{m}} }.\label{eqn:SIR}
\end{align}
Note that the distribution of ${\rm SIR}_{m,n,0}$ is affected by $\mathbf T$.

We assume that file $n$ delivered from tier $m$ can be decoded correctly at $u_0$ if ${\rm SIR_{m,n,0}\geq\tau_{m}}$, where $\tau_{m}$ represents a threshold for tier $m$~\cite{wen2016cache}.
Requesters are mostly concerned about whether their desired files can be successfully received. 
In the following, we introduce the performance metrics in the three cases of file popularity distribution. \\
\textbf{Perfect file popularity distribution:} When the exact value of $\mathbf a$ is known, we adopt the probability that a randomly requested file by $u_0$ is successfully transmitted, called the STP~\cite{Wang2017Joint}:
\begin{align}
q\left(\mathbf a,\mathbf T\right)\triangleq& \sum_{m\in\mathcal M}\sum_{n\in\mathcal N} a_n A_{m,n}\left(\mathbf T\right){\rm Pr}\left[{\rm SIR}_{m,n,0}\geq\tau_{m}\right]\nonumber\\
=&\sum_{m\in\mathcal M}\sum_{n\in\mathcal N}{\frac{a_n T_{m,n}}{\sum_{l\in\mathcal{M}}{\theta_{l,m}T_{l,n}}+\eta_{m}}}, \label{eqn:STP}
\end{align}
as the performance metric,\footnote{
\blue{Note that the STP with fixed STR thresholds can be interpreted as the STP under multicast at the high user density region~\cite{cui2016analysis,cui2017analysis,Wang2017Joint}, and is a widely used performance metric in wireless caching~\cite{cui2016analysis,cui2017analysis,li2016optimization,wen2016cache,Wang2017Joint}, as it is tractable and is also closely related to some other important performance metrics, such as average transmission rate and average transmission delay.}}
where $\theta_{l,m}$ and $\eta_{m}$ are given by: 
\begin{align}
\theta_{l,m}=\frac{2\lambda_{l}}{\alpha\lambda_m}\left(\frac{P_l}{P_m}\tau_{m} \right)^{\frac{2}{\alpha}}\left(B'\left(\frac{2}{\alpha},1-\frac{2}{\alpha}, \frac{1}{1+\tau_{m}} \right)-B\left(\frac{2}{\alpha},1-\frac{2}{\alpha}\right)\right)+\frac{\lambda_{l}}{\lambda_m}\left(\frac{P_l}{P_m}\right)^{\frac{2}{\alpha}},\label{eqn:theta}
\end{align}
\begin{equation}
\eta_{m}=\sum_{l\in\mathcal{M}}\frac{2\lambda_l}{\alpha\lambda_m}\left(\frac{P_l}{P_m}\tau_{m}\right)^{\frac{2}{\alpha}}B\left(\frac{2}{\alpha},1-\frac{2}{\alpha}\right),\label{eqn:eta}
\end{equation}
respectively. 
\blue{Here, $B'(x,y,z)\triangleq\int_z^1 u^{x-1}(1-u)^{y-1}\mathrm{d}u$ and $B(x,y)\triangleq\int_0^1 u^{x-1}(1-u)^{y-1}\mathrm{d}u$ denote the complementary incomplete Beta function and the Beta function, respectively.}
Note that $q(\mathbf a,\mathbf T)$ is a linear function of $\mathbf a$ and a nonconcave function of $\mathbf T$.\\
\textbf{Imperfect file popularity distribution:} When the exact value of $\mathbf a$ is not known except that it falls within a known set $\mathcal A$, we adopt the worst-case STP: 
\begin{align}
q_{\text{wt}}(\mathcal{A},\mathbf{T})\triangleq \min_{\mathbf a\in\mathcal{A}} q(\mathbf{a},\mathbf{T}),\label{eqn:wtSTP}
\end{align}
as the performance metric.\footnote{Note that one of the major techniques for designing systems that are robust against modeling uncertainties is to optimize the worst-case performance.}\\
\textbf{Unknown file popularity distribution:} When there is no prior information of $\mathbf{a}$, but $\{\pi_s(t):s\in\mathcal U\}$ can be obtained at each slot $t$ for some $\mathcal U\subseteq\Phi_u$, we adopt the STP in the stochastic form, called the stochastic STP:
\begin{align}
q_{\text{st}}(\mathbf a,\mathbf T)\triangleq\mathbb{E}\left[q(\boldsymbol\xi,\mathbf T)\right], \label{eqn:stSTP}
\end{align}
as the performance metric, where \blue{$\xi_n\triangleq\mathbf I[\pi_0(t)=n|\pi_0(t)\neq0]$} with $\mathbf I[\cdot]$ denoting the indicator function, $\boldsymbol\xi\triangleq(\xi_n)_{n\in\mathcal N}$, \blue{and the expectation is take over $\boldsymbol\xi$.}
Note that $q_{\text{st}}(\mathbf a,\mathbf T)=q(\mathbf{a},\mathbf{T})$ as \blue{$\mathbb{E}\left[\mathbf I[\pi_0(t)=n|\pi_0(t)\neq0]\right]={\rm Pr}[\pi_0(t)=n|\pi_0(t)\neq0]$} for all $n\in\mathcal N$ and $t\in\{1,2,\dots\}$.

\begin{figure}[t]
\begin{center}
{\resizebox{12cm}{!}{\includegraphics{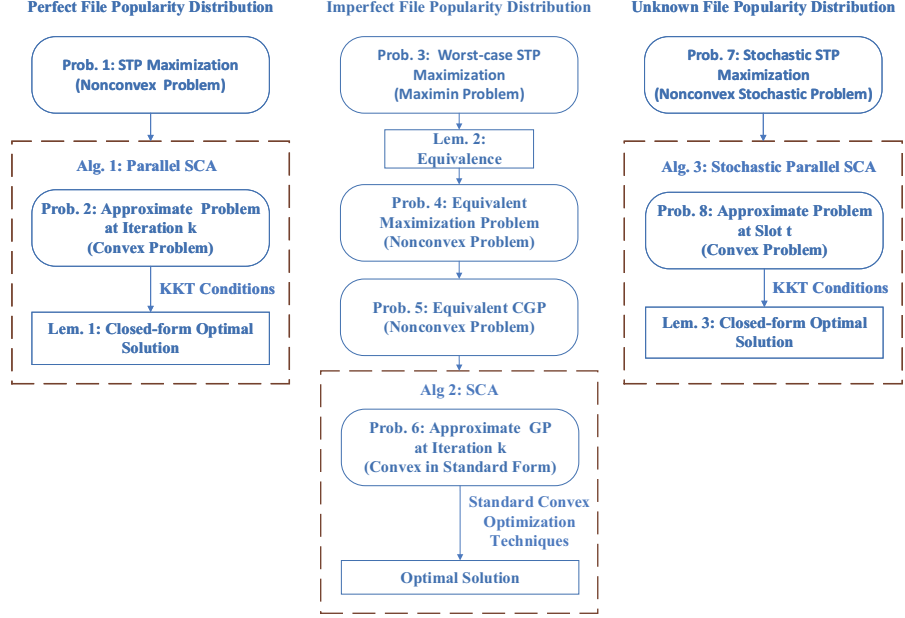}}}
\end{center}
\vspace{-3mm}
\caption{\small{The optimization problems and corresponding solutions in the cases of perfect, imperfect and unknown file popularity distribution.}}
\vspace{-3mm}
\label{fig:problem}
\end{figure}

In Section~\ref{Sec:pe}, Section~\ref{Sec:ro} and Section~\ref{Sec:sto}, we shall maximize the STP, the worst-case STP and the stochastic STP in the cases of perfect, imperfect and unknown file popularity distribution, respectively, as shown in Fig~\ref{fig:problem}.\footnote{
\blue{Based on the solutions in this paper. we can obtain promising caching designs under multicast at the general usesr density region using the method proposed in our previous work~\cite{cui2016analysis,cui2017analysis}.}}

\section{Performance Optimization for Perfect File Popularity Distribution}\label{Sec:pe}

In this section, we consider the case of perfect file popularity distribution. In this case, we would like to maximize the STP, by optimizing the caching probabilities. We formulate the optimal random caching design problem as follows.

\begin{Prob}[Optimization for Perfect File Popularity Distribution]\label{prob:rand}
	\begin{align}
	q^*(\mathbf a) \triangleq \max_{\mathbf{T}}\quad  &q\left(\mathbf a,\mathbf{T}\right)\nonumber\\
	\text{s.t.}  \quad &\eqref{eqn:T1},\eqref{eqn:T2},\nonumber
	\end{align}
	where $ q\left(\mathbf a,\mathbf{T}\right) $ is given by \eqref{eqn:STP}.
\end{Prob}

Problem~\ref{prob:rand} is equivalent to Problem~0 in \cite{wen2016cache}. It is non-convex (as the objective function is non-convex in $\mathbf T$), and in \cite{wen2016cache} a suboptimal solution of it is obtained by solving an approximate convex problem. 
In the following, we extend the technique in \cite{Wang2017Joint} for the case of $M=2$ to the case of $M\geq2$, and develop an efficient parallel iterative algorithm to obtain a stationary point of Problem~\ref{prob:rand} using parallel SCA. 
Different from the cyclic computation mechanism in \cite{Wang2017Joint}, the parallel computation mechanism here can speed up the computation, especially for large $M$. Specifically, this algorithm updates the caching probabilities of the $M$ tiers, i.e., $\mathbf{T}_m\triangleq(T_{m,n})_{n\in\mathcal{N}}$, $m\in\mathcal M$, at each iteration in a parallel manner, by maximizing $M$ approximate functions of $q\left(\mathbf a,\mathbf{T}\right)$.


For notation convenience, define\blue{:}
\begin{align}
q_{j}\left(\mathbf a,\mathbf{T}_m,\mathbf{T}_{-m}\right)\triangleq\sum_{n\in\mathcal{N}}{\frac{a_n T_{j,n}}{\sum_{l\in\mathcal{M}}{\theta_{l,j}T_{l,n}}+\eta_{j}}},\label{eqn:qm}
\end{align}
where $\mathbf T_{-m}\triangleq\left(\mathbf T_j\right)_{j\in\mathcal M,j\neq m}$. Note that $q(\mathbf a,\mathbf T)$ can be rewritten as:
\begin{align}
q\left(\mathbf a,\mathbf{T}_m,\mathbf{T}_{-m}\right)\triangleq\sum_{j\in\mathcal M}q_{j}\left(\mathbf a,\mathbf{T}_m,\mathbf{T}_{-m}\right).\nonumber
\end{align}
Let $\mathbf{T}_m^{(k)}$ denote the caching probabilities of tier $m$ obtained at iteration $k$, and denote $\mathbf{T}^{(k)}\triangleq\big(\mathbf{T}_m^{(k)}\big)_{m\in\mathbf{M}}$. 
At iteration $k+1$, choose\blue{:} 
\begin{align}
&h_m\big(\mathbf{a},\mathbf{T}_m,\mathbf{T}^{(k)}\big)\triangleq\nonumber\\
&q_{m}\left(\mathbf{a},\mathbf{T}_m,\mathbf{T}_{-m}^{(k)}\right)+\sum_{j\in\mathcal{M},j\neq m}\left(q_{j}\left(\mathbf{a},\mathbf{T}_m^{(k)},\mathbf{T}_{-m}^{(k)}\right)
-\sum_{n\in\mathcal{N}}\frac{a_n\theta_{m,j}T_{j,n}^{(k)}\left(T_{m,n}-T_{m,n}^{(k)}\right)}{\left({\sum_{l\in\mathcal{M}}{\theta_{l,j}T_{l,n}^{(k)}}+\eta_{j}}\right)^2}\right), \label{eqn:randap}
\end{align}
as an approximation function of $q(\mathbf a,\mathbf{T})$ for updating $\mathbf T_m$.
Note that the strongly concave component function of $q\big(\mathbf{a},\mathbf{T}_m,\mathbf{T}_{-m}^{(k)}\big)$, i.e., $q_{m}\big(\mathbf{a},\mathbf{T}_m,\mathbf{T}_{-m}^{(k)}\big)$, is left unchanged, and the other nonconcave (actually convex) component functions, i.e., $q_{j}\big(\mathbf{a},\mathbf{T}_m,\mathbf{T}_{-m}^{(k)}\big)$, $j\in\mathcal{M}$, $j\neq m$, are linearized at $\mathbf{T}_m=\mathbf{T}_m^{(k)}$. This choice of the approximate function is beneficial from several aspects\cite{Wang2017Joint}. Firstly, it can guarantee the convergence of the algorithm to a stationary point of Problem~\ref{prob:rand}, which will be shown in Theorem~\ref{thm:conv-rand}. Secondly, it usually leads to fast convergence of the algorithm by exploiting the partial concavity of the objective function, which will be shown in Fig.~\ref{fig:simulation-convergence}. Thirdly, it yields a closed-form optimal solution of the optimization problem for each tier at each iteration, which will be shown in Lemma~\ref{lem:closed-formrand}, and hence a low-complexity algorithm.

Specifically, at iteration $k$, we first solve the following problem for each tier $m\in\mathcal M$ separately, in a parallel manner.
\begin{Prob}[Approximate Convex Problem of Problem~\ref{prob:rand} for Tier $m$ at Iteration $k$]\label{prob:randsp}
\begin{align}
\overline{\mathbf{T}}_m^{(k)} \triangleq \mathop{\arg\max}_{\mathbf{T}_m}\ &h_m\big(,\mathbf{a},\mathbf{T}_m,\mathbf{T}^{(k-1)}\big) \nonumber\\
\text{s.t.}  \quad &0\leq T_{m,n}\leq 1,\quad n\in\mathcal N, \label{eqn:Tm1}\\
&\sum\limits_{n\in\mathcal N}T_{m,n}=K_m. \label{eqn:Tm2}
\end{align}
\end{Prob}

Problem~\ref{prob:randsp} is a convex optimization problem and Slater's condition is satisfied, implying that strong duality holds. Based on KKT conditions, we can obtain \blue{a} closed-form optimal solution of Problem~\ref{prob:randsp}.
\begin{Lem}[Optimal Solution of Problem~\ref{prob:randsp}]\label{lem:closed-formrand}
For all $m\in\mathcal{M}$, the optimal solution of Problem~\ref{prob:randsp} is given by\blue{:} 
\begin{align}
&\overline T_{m,n}^{(k)}=
\left[\frac{1}{\theta_{m,m}}\sqrt{\frac{a_n\left(\sum\limits_{l\in\mathcal{M},l\neq m}\theta_{l,m}T_{l,n}^{(k-1)}+\eta_{m}\right)}{ \nu_m^{*({k})} + \sum\limits_{j\in\mathcal{M},j\neq m}\frac{a_n\theta_{m,j}T_{j,n}^{(k-1)}}{\left({\sum\limits_{l\in\mathcal{M}}{\theta_{l,j}T_{l,n}^{(k-1)}}+\eta_{j}}\right)^2} }} - \frac{\sum\limits_{l\in\mathcal{M},l\neq m}\theta_{l,m}T_{l,n}^{(k-1)}+\eta_{m}}{\theta_{m,m}}\right]^1_0, \label{eqn:BSUM-opt-sol}
\end{align}
where $[x]^1_0\triangleq\min\left\{\max\{x,0\},1\right\}$ and $\nu_m^{*({k})}$ is the Lagrange multiplier that satisfies $\sum_{n\in\mathcal{N}}\overline T_{m,n}^{(k)}=K_m$.
\end{Lem} 

\blue{Note that $\nu_m^{*({k})}$ in Lemma~\ref{lem:closed-formrand} can be efficiently obtained by using bisection search which achieves a desired accuracy $\epsilon$ with computational complexity $\mathcal O\left(M^2N\log(1/\epsilon)\right)$.}
Then, we update the caching probabilities of tier $m$ by:
\begin{align}
&\mathbf{T}_m^{(k)}=(1-\gamma^{(k)})\mathbf{T}_m^{(k-1)}+\gamma^{(k)}\overline{\mathbf{T}}_m^{(k)},\label{eqn:updateTm}
\end{align}
where $\gamma^{(k)}$ is a positive diminishing stepsize satisfying 
\begin{align}
&\gamma^{(k)}>0,\quad \lim_{k\to\infty}\gamma^{(k)}=0,\quad \sum_{k=1}^\infty\gamma^{(k)}=\infty,\quad \sum_{k=1}^\infty\left(\gamma^{(k)}\right)^2<\infty. \label{eqn:gamma}
\end{align}
Finally, the details of the proposed parallel iterative algorithm are summarized in Algorithm~\ref{alg:rand}. Based on~\cite[Theorem~1]{razaviyayn2014parallel}, we can show the following result.

\begin{Thm}[Convergence of Algorithm~\ref{alg:rand}]\label{thm:conv-rand} 
If the stepsize $\{\gamma^{(k)}\}$ satisfies \eqref{eqn:gamma}, then every limit point of $\{\mathbf{T}^{(k)}\}$ generated by Algorithm~\ref{alg:rand} is a stationary point of Problem 1.

\end{Thm}
\begin{IEEEproof}
Please refer to Appendix A.
\end{IEEEproof}

\begin{algorithm}[t]
    \caption{Obtaining A Stationary Point of Problem~\ref{prob:rand}}
\begin{small}
        \begin{algorithmic}[1]
           \STATE \textbf{initialization}: choose any feasible solution $\mathbf{T}^{(0)}$ of Problem~\ref{prob:rand} as the initial point, and set $k=1$.\\
           \STATE \textbf{repeat}
           \STATE \quad for all $m\in\mathcal{M}$, compute $\overline{\mathbf{T}}_m^{(k)}$ according to \eqref{eqn:BSUM-opt-sol}, and update $\mathbf{T}_m^{(k)}$ according to \eqref{eqn:updateTm}.
           \STATE\quad set $k=k+1$.
           \STATE \textbf{until} some convergence criteria is met.
    \end{algorithmic}\label{alg:rand}
    \end{small}
\end{algorithm}

\section{Robust Optimization for Imperfect File Popularity Distribution}\label{Sec:ro}

In this section, we consider the case of imperfect file popularity distribution. In this case, we would like to maximize the worst-case STP, by optimizing the caching probabilities. We formulate the robust optimal random caching design problem 
as follows.

\begin{Prob}[Robust Optimization for Imperfect File Popularity Distribution]\label{prob:ro}
\begin{align}
q_{\text{wt}}^* \triangleq \max_{\mathbf{T}}\ &\underbrace{\min_{\mathbf a\in\mathcal{A}} q(\mathbf{a},\mathbf{T})}_{= q_{\text{wt}}(\mathcal{A},\mathbf{T})}\nonumber\\
\text{s.t.}  \quad &\eqref{eqn:T1},\eqref{eqn:T2},\nonumber
\end{align}
where $q(\mathbf{a},\mathbf{T})$ is given by \eqref{eqn:STP}. 
\end{Prob}

Problem~\ref{prob:ro} is a challenging maximin problem, which does not lie in the category of convex-concave games that can be easily solved (as $q(\mathbf a,\mathbf T)$ is a nonconcave function of $\mathbf T$). In the following, we solve it in two steps.

Firstly, we transform the maximin problem in Problem~\ref{prob:ro} to an equivalent maximization problem. As the inner problem $\min_{\mathbf{a}\in\mathcal{A}}q\left(\mathbf a,\mathbf{T}\right)$ is a linear programming (LP) with respect to $\mathbf{a}$ and strong duality holds for LP, the inner problem $\min_{\mathbf{a}\in\mathcal{A}}q\left(\mathbf a,\mathbf{T}\right)$ shares the same optimal value with its dual problem. Thus, we can transform Problem~\ref{prob:ro} to the following equivalent maximization problem by replacing the inner problem with its dual problem.

\begin{Prob}[Equivalent Problem of Problem~\ref{prob:ro}]\label{prob:dualmax}
	\begin{align}
q_{\text{wt}}^{\star} \triangleq \max_{\mathbf{T},\boldsymbol{\lambda}\succeq\mathbf0,\boldsymbol{\mu}\succeq\mathbf0,{\nu}} \quad &\sum_{n\in\mathcal{N}}{(\lambda_n \underline a_n-\mu_n \overline a_n)}-\nu\nonumber\\
	\text{s.t.} \quad &\eqref{eqn:T1},\eqref{eqn:T2},\nonumber\\
&\sum_{m\in\mathcal{M}}{\frac{T_{m,n}}{\sum_{l\in\mathcal{M}}{\theta_{l,m}T_{l,n}}+\eta_{m}}}+\mu_n-\lambda_n+\nu=0,\quad n\in\mathcal{N},\label{eqn:dual1}
	\end{align}
where $\boldsymbol{\lambda}\triangleq(\lambda_n)_{n\in\mathcal{N}}$ and $\boldsymbol{\mu}\triangleq(\mu_n)_{n\in\mathcal{N}}$. 
\end{Prob}

Note that $\boldsymbol{\lambda}$, $\boldsymbol{\mu}$ and $\nu$ are dual variables for the dual problem of the inner problem, corresponding to $a_n\geq\underline a_n$, $a_n\leq\overline a_n$, $n\in\mathcal N$, and $\sum_{n\in\mathcal N}a_n=1$, respectively.

\begin{Lem}[Equivalence between Problem~\ref{prob:ro} and Problem~\ref{prob:dualmax}]\label{lem:roeq}
Problem~\ref{prob:ro} and Problem~\ref{prob:dualmax} have the same optimal value and optimal caching probabilities.
\end{Lem}
\begin{IEEEproof}
Please refer to Appendix B.
\end{IEEEproof}

Based on Lemma~\ref{lem:roeq}, we can solve Problem~\ref{prob:dualmax} instead of Problem~\ref{prob:ro}. Problem~\ref{prob:dualmax} is nonconvex, as the constraints in \eqref{eqn:dual1} are nonconvex.
In what follows, we show how to obtain a stationary point of Problem~\ref{prob:dualmax} using SCA.
We first rewrite $\nu$ as $\nu_1-\nu_2$ with $\nu_1,\nu_2>0$, and define new \blue{variables} $\mathbf{x}\triangleq(x_{m,n})_{m\in\mathcal{M},n\in\mathcal{N}}$:\footnote{Note that $\eta_{m}>0$, as $B\left(\frac{2}{\alpha},1-\frac{2}{\alpha}\right)>0$, and in most practical cases, $\theta_{l,m}>0$~\cite{Wang2017Joint}. Thus, in the rest of this paper, we consider the case where $x_{m,n}>0$ for all $m\in\mathcal{M}$ and $n\in\mathcal{N}$.}
\begin{align}
 x_{m,n}=\sum_{l=1}^M \theta_{l,m} T_{l,n}+ \eta_{m}, \quad m\in\mathcal{M}, n\in\mathcal{N}. \label{eqn:xeq}
\end{align}
We also introduce a new variable $y>0$ which serves as a lower bound of the objective function of Problem~\ref{prob:dualmax}:
\begin{align}
y\leq\sum_{n\in\mathcal{N}}{(\lambda_n \underline a_n-\mu_n \overline a_n)}-\nu. \label{eqn:yneq}
\end{align}
Therefore, Problem~\ref{prob:dualmax} can be equivalently transformed to the following problem.\footnote{For ease of analysis, in Problem~\ref{prob:roep}, we consider $\mathbf T\succ0$ instead of $\mathbf T\succeq0$, which does not change the optimal value or affect the numerical solution.}

\begin{Prob}[Equivalent Problem of Problem~\ref{prob:dualmax}]\label{prob:roep}
\begin{align}
\max_{\substack{\mathbf{T},\mathbf{x},\boldsymbol{\lambda},\boldsymbol{\mu}\succ\mathbf 0\\ {\nu_1},{\nu_2},y>0}} \quad &y \nonumber\\
\text{s.t.} \quad 
&\frac{y+\sum_{n\in\mathcal{N}}{\mu_n \overline a_n}+\nu_1}{\sum_{n\in\mathcal{N}}{\lambda_n \underline a_n}+\nu_2}\leq1, \label{eqn:c1}\\
&\frac{\lambda_n+\nu_2}{\sum_{m\in\mathcal{M}} {T_{m,n}}{x_{m,n}^{-1}}+\mu_n+\nu_1}\leq1, \quad n\in\mathcal{N},\label{eqn:c2}\\
&\frac{\sum_{l\in\mathcal{M}} \theta_{l,m} T_{l,n} + \eta_{m}}{x_{m,n}}\leq1, 
\quad m\in\mathcal{M},\ n\in\mathcal{N}, \label{eqn:x1}\\
&T_{m,n}\leq 1,\quad m\in\mathcal M,\ n\in\mathcal N, \label{eqn:T3}\\
&\sum\limits_{n\in\mathcal N}T_{m,n}\leq K_m,\quad m\in\mathcal M. \label{eqn:T4}
\end{align}
\end{Prob}

Note that the inequality constraints in \eqref{eqn:c2}, \eqref{eqn:x1} and \eqref{eqn:T4} are active at any optimal solution of Problem~\ref{prob:roep}, and hence can replace the equality constraints in \eqref{eqn:dual1}, \eqref{eqn:xeq} and \eqref{eqn:T2}, respectively.
In Problem~\ref{prob:roep}, a monomial \blue{is maximized} subject to upper bounds on posynomials (i.e., \eqref{eqn:x1}, \eqref{eqn:T3} and \eqref{eqn:T4}) and upper bounds on the ratios of posynomials (i.e., \eqref{eqn:c1} and \eqref{eqn:c2}). Thus,  Problem~\ref{prob:roep} is a CGP, and can be solved by the method proposed in \cite{4275017}, which is based on SCA.
The main idea is to solve a sequence of successively refined geometric programmings (GPs), each of which is obtained by approximating the denominators of the ratios of posynomials in \eqref{eqn:c1} and \eqref{eqn:c2} with monomials. 
Specifically, at iteration $k$, update $\big(\mathbf{T}^{(k)},\mathbf{x}^{(k)},\boldsymbol{\lambda}^{(k)},\boldsymbol{\mu}^{(k)},{\nu_1}^{(k)},{\nu_2}^{(k)},y^{(k)}\big)$ by solving the following approximate GP of Problem~\ref{prob:roep}, which is parameterized by $\big(\mathbf{T}^{(k-1)},\mathbf{x}^{(k-1)},\boldsymbol{\lambda}^{(k-1)},\boldsymbol{\mu}^{(k-1)},{\nu_1}^{(k-1)},{\nu_2}^{(k-1)}\big)$ obtained at iteration $k-1$.

\begin{Prob}[Approximate GP at Iteration $k$]\label{prob:roap}
	\begin{align}
&\hspace{-32mm}\big(\mathbf{T}^{(k)},\mathbf{x}^{(k)},\boldsymbol{\lambda}^{(k)},\boldsymbol{\mu}^{(k)},{\nu_1}^{(k)},{\nu_2}^{(k)},y^{(k)}\big)\triangleq\mathop{\arg\max}_{\substack{\mathbf{T},\mathbf{x},\boldsymbol{\lambda},\boldsymbol{\mu}\succ\mathbf 0\\ {\nu_1},{\nu_2},y>0}} \quad y \nonumber\\
	\text{s.t.} \quad &\eqref{eqn:x1},\eqref{eqn:T3},\eqref{eqn:T4},\nonumber\\
&\frac{y+\sum_{n=1}^N{\mu_n \overline a_n}+\nu_1}{\prod_{n\in\mathcal{N}}{\left(\frac{\lambda_n \underline a_n}{\sigma_n^{(k)}}\right)^{\sigma_n^{(k)}}}\left(\frac{\nu_2}{\gamma_1^{(k)}}\right)^{\gamma_1^{(k)}}}\leq1, \\
&\frac{\lambda_n+\nu_2}{\prod_{m\in\mathcal{M}} \left(\frac{T_{m,n}x_{m,n}^{-1}}{\beta_{m,n}^{(k)}}\right)^{\beta_{m,n}^{(k)}}\left(\frac{\mu_n}{\gamma_{2,n}^{(k)}}\right)^{\gamma_{2,n}^{(k)}}\left(\frac{\nu_1}{\gamma_{3,n}^{(k)}}\right)^{\gamma_{3,n}^{(k)}}}\leq1,\quad n\in\mathcal{N},
	\end{align}
where
\begin{align}
\sigma_{n}^{(k)}\triangleq\ &\frac{\lambda_n^{(k-1)} \underline a_n}{\sum_{n\in\mathcal{N}}{\lambda_n^{(k-1)} \underline a_n}+\nu_2^{(k-1)}},\nonumber\\
\beta_{m,n}^{(k)}\triangleq\ &\frac{T_{m,n}^{(k-1)}\left(x_{m,n}^{(k-1)}\right)^{-1}}{\sum_{m\in\mathcal{M}} {T_{m,n}^{(k-1)}}{\left(x_{m,n}^{(k-1)}\right)^{-1}}+\mu_n^{(k-1)}+\nu_1^{(k-1)}},\nonumber\\
\gamma_{1}^{(k)}\triangleq\ &\frac{\nu_2^{(k-1)}}{\sum_{n\in\mathcal{N}}{\lambda_n^{(k-1)} \underline a_n}+\nu_2^{(k-1)}},\nonumber\\
\gamma_{2,n}^{(k)}\triangleq\ &\frac{\mu_n^{(k-1)}}{\sum_{m\in\mathcal{M}} {T_{m,n}^{(k-1)}}{\left(x_{m,n}^{(k-1)}\right)^{-1}}+\mu_n^{(k-1)}+\nu_1^{(k-1)}},\nonumber\\
\gamma_{3,n}^{(k)}\triangleq\ &\frac{\nu_1^{(k-1)}}{\sum_{m\in\mathcal{M}} {T_{m,n}^{(k-1)}}{\left(x_{m,n}^{(k-1)}\right)^{-1}}+\mu_n^{(k-1)}+\nu_1^{(k-1)}}.\nonumber
\end{align}
\end{Prob}

Problem \ref{prob:roap} is a standard GP, which can be readily transformed into a convex problem and solved \blue{using standard convex optimization techniques, such as the barrier method which achieves a desired accuracy $\epsilon$ with computational complexity $\mathcal{O}(M^3N^3\log(MN/\epsilon))$.} The details for solving Problem~\ref{prob:roep} are summarized in Algorithm~\ref{alg:ro}. 
By the convergence result in~\cite[Proposition~3]{4275017}, and by comparing the KKT conditions of Problem~\ref{prob:dualmax} and Problem~\ref{prob:roep}, we have the following result.
\begin{Thm}[Convergence of Algorithm~\ref{alg:ro}]\label{thm:conv-ro}
$\left(\mathbf{T}^{(k)}\right.$, $\mathbf{x}^{(k)}$, $\boldsymbol{\lambda}^{(k)}$, $\boldsymbol{\mu}^{(k)}$, ${\nu_1}^{(k)}$, ${\nu_2}^{(k)}$, $\left.y^{(k)}\right)$ obtained by Algorithm~\ref{alg:ro} converges to a stationary point of Problem~\ref{prob:roep}, as $k\to \infty$. Furthermore, the limit point of $\left\{\left(\mathbf{T}^{(k)}\right.\right.$, $\boldsymbol{\lambda}^{(k)}$, $\boldsymbol{\mu}^{(k)}$, $\left.\left.{\nu_1}^{(k)}-{\nu_2}^{(k)}\right)\right\}$ is a stationary point of Problem~\ref{prob:dualmax}.
\end{Thm}
\begin{IEEEproof}
Please refer to Appendix C.
\end{IEEEproof}

\begin{algorithm}[t]
    \caption{Obtaining A Stationary Point of Problem~\ref{prob:roep}}
\begin{small}
        \begin{algorithmic}[1]
           \STATE \textbf{initialization}: choose any feasible solution $\big(\mathbf{T}^{(0)}$, $\mathbf{x}^{(0)}$, $\boldsymbol{\lambda}^{(0)}$, $\boldsymbol{\mu}^{(0)}$, ${\nu_1}^{(0)}$, ${\nu_2}^{(0)}$, $y^{(0)}\big)$ of Problem~\ref{prob:roep} as the initial point, and set $k=1$.\\
           \STATE \textbf{repeat}
           \STATE \quad compute $\big(\mathbf{T}^{(k)}$, $\mathbf{x}^{(k)}$, $\boldsymbol{\lambda}^{(k)}$, $\boldsymbol{\mu}^{(k)}$, ${\nu_1}^{(k)}$, ${\nu_2}^{(k)}$, $y^{(k)}\big)$ by transforming Problem~\ref{prob:roap} into a GP\\ \quad in convex form, and solving it with standard convex optimization techniques.
           \STATE\quad set $k=k+1$.
           \STATE \textbf{until} some convergence criteria is met.
    \end{algorithmic}\label{alg:ro}
    \end{small}
\end{algorithm}

\section{Stochastic Optimization for Unknown File Popularity Distribution}\label{Sec:sto}
In this section, we consider the case of unknown file popularity distribution. In this case, we would like to maximize the stochastic STP, by optimizing the caching probabilities. We formulate the stochastic optimal random caching design problem as follows.

\begin{Prob}[Stochastic Optimization for Unknown File Popularity Distribution]\label{prob:sto}
	\begin{align}
\max_{\mathbf{T}}\quad  &\underbrace{\mathbb{E}\left[q(\boldsymbol\xi,\mathbf T)\right]}_{=q_\text{st}(\mathbf a,\mathbf{T})}\nonumber\\
	\text{s.t.}  \quad &\eqref{eqn:T1},\eqref{eqn:T2},\nonumber
	\end{align}
where $q\left(\mathbf a,\mathbf T\right)$ is given by \eqref{eqn:STP}.
\end{Prob}

\blue{Note that although $\mathbb{E}\left[q(\boldsymbol\xi,\mathbf T)\right]$ cannot be calculated without knowledge of the statistics of $\boldsymbol\xi$, it can be optimized using stochastic optimization.\footnote{\blue{The basic idea of stochastic optimization is to optimize a function in the presence of randomness based on the fact that realizations of random parameters can be obtained.}}}
Problem~\ref{prob:sto} is a nonconvex stochastic optimization problem, which is more challenging than a convex one.
In the following, we develop an efficient parallel iterative algorithm to obtain a stationary point of Problem~\ref{prob:sto}, using stochastic parallel SCA~\cite{7412752}. Similarly, the parallel computation mechanism here can speed up the computation, especially for large $M$. Specifically, this algorithm updates the caching probabilities of the $M$ tiers, i.e., $\mathbf{T}_m\triangleq(T_{m,n})_{n\in\mathcal{N}}$, $m\in\mathcal M$, at each slot $t$ in a parallel manner, by maximizing $M$ approximate functions of $q_{st}(\mathbf a,\mathbf{T})$.

Let $\mathbf{T}_m^{(t)}$ denote the caching probabilities of tier $m$ obtained at slot $t$, and denote $\mathbf{T}^{(t)}\triangleq\big(\mathbf{T}_m^{(t)}\big)_{m\in\mathbf{M}}$. 
At slot $t$, choose\blue{:}
\begin{align}
&\widehat{h}_{m}\left(\boldsymbol\xi^{(t)},\mathbf{T}_m,\mathbf{T}^{(t-1)}\right)=\rho^{(t)}\Bigg(q_{m}\left(\boldsymbol\xi^{(t)},\mathbf{T}_m,\mathbf{T}_{-m}^{(t-1)}\right)-\nonumber\\
&\sum_{j\in\mathcal{M},j\neq m}\sum_{n\in\mathcal{N}}\frac{\xi^{(t)}_n\theta_{m,j,K_j}T_{j,n}^{(t-1)}\left(T_{m,n}-T_{m,n}^{(t-1)}\right)}{\left({\sum_{l\in\mathcal{M}}{\theta_{l,j,K_j}T_{l,n}^{(t-1)}}+\eta_{j,K_j}}\right)^2}\Bigg)
+(1-\rho^{(t)})\sum_{n\in\mathcal N}\left({T}_{m,n}-{T}_{m,n}^{(t-1)}\right){{f}^{(t-1)}_{m,n}}\label{eqn:hhat}
\end{align}
as an approximation function of $q_{st}(\mathbf a,\mathbf{T})$ for updating $\mathbf T_m$.
Here, \blue{$\xi^{(t)}_n\triangleq\frac{\sum_{s\in\mathcal U}\mathbf I\left[\pi^{(t)}_s=n\right]}{\sum_{s\in\mathcal U}\mathbf I\left[\pi^{(t)}_s\neq0\right]}$,}\footnote{
\blue{We consider a simple way of making use of instantaneous file requests at each slot without assuming any apriori information of the file popularity distribution. Our focus here is to optimize random caching design using stochastic optimization, instead of pure estimation of file popularity.}}
$\boldsymbol\xi^{(t)}\triangleq(\xi^{(t)}_n)_{n\in\mathcal N}$, 
$\rho^{(t)}$ is a positive diminishing stepsize satisfying\blue{:}
\begin{align}
\rho^{(t)}>0,\quad \lim_{t\to\infty}\rho^{(t)}=0,\quad \sum_{t=1}^\infty\rho^{(t)}=\infty,\quad \sum_{t=1}^\infty\left(\rho^{(t)}\right)^2<\infty,\label{eqn:rho}
\end{align}
and $f^{(t)}_{m,n}$ is given by\blue{:}
\begin{align}
&f^{(t)}_{m,n}=(1-\rho^{(t)})f^{(t-1)}_{m,n}+\rho^{(t)}\left(\frac{\xi_n}{{\sum_{l\in\mathcal{M}}{\theta_{l,m}T_{l,n}^{(t)}}+\eta_{m}}}-\sum_{j\in\mathcal{M}}\frac{\xi_n\theta_{m,j}T_{j,n}^{(t)}}{\left({\sum_{l\in\mathcal{M}}{\theta_{l,j}T_{l,n}^{(t)}}+\eta_{j}}\right)^2}\right),\label{eqn:ft}
\end{align}
where $f^{(0)}_{m,n}=0$ , $m\in\mathcal M$, $n\in\mathcal N$.

Note that the strongly concave component function of $q\big(\boldsymbol\xi^{(t)},\mathbf{T}_m,\mathbf{T}_{-m}^{(t-1)}\big)$, i.e., $q_{m}\big(\boldsymbol\xi^{(t)},\mathbf{T}_m,$
$\mathbf{T}_{-m}^{(t-1)}\big)$, is left unchanged, and the other nonconcave (actually convex) component functions, i.e., $q_{j}\big(\boldsymbol\xi^{(t)},\mathbf{T}_m,\mathbf{T}_{-m}^{(t-1)}\big)$, $j\in\mathcal{M}$, $j\neq m$, are linearized at $\mathbf{T}_m=\mathbf{T}_m^{(t-1)}$.
In addition, note that the \blue{approximation} of $a_n$ at each slot $t$, i.e., $\xi^{(t)}_n$, becomes more accurate as ${\sum_{s\in\mathcal U}\mathbf I\left[\pi^{(t)}_s\neq0\right]}$ increases, and
the \blue{approximation} of $\nabla_{{T}_{m,n}}q_\text{st}\big(\mathbf a,\mathbf{T}^{(t)}\big)$ based on accumulated instantaneous file requests $\{\pi_s(t):s\in\mathcal U\}$, $t\in\{1,2,\dots,t-1\}$, i.e., ${f}^{(t)}_{m,n}$, becomes more accurate as $t$ increases.
This choice of the approximate function, $\widehat{h}_{m}\left(\boldsymbol\xi^{(t)},\mathbf{T}_m,\mathbf{T}^{(t-1)}\right)$, given in \eqref{eqn:hhat}, is beneficial for similar reasons as in the case of perfect file popularity distribution.


Specifically, at slot $t$, we first solve the following problem for each tier $m\in\mathcal M$ separately, in a parallel manner.
\begin{Prob}[Approximate Convex Problem of Problem~\ref{prob:sto} for Tier $m$ at Slot $t$]\label{prob:stoap}
\begin{align}
\widehat{\mathbf{T}}_m^{(t)}\triangleq\mathop{\arg\max}_{\mathbf{T}_m} \quad &\widehat{h}_{m}\left(\boldsymbol\xi^{(t)},\mathbf{T}_m,\mathbf{T}^{(t-1)}\right) \nonumber\\*
\text{s.t.}  \quad &\eqref{eqn:Tm1},\eqref{eqn:Tm2} \nonumber
\end{align}
\end{Prob}

Problem~\ref{prob:stoap} is a convex optimization problem and Slater's condition is satisfied, implying that strong duality holds. Based on KKT conditions, we can obtain \blue{a} closed-form optimal solution of Problem~\ref{prob:stoap}. 

\begin{Lem}[Optimal Solution of Problem~\ref{prob:stoap}]\label{lem:closed-formsto}
For all $m\in\mathcal{M}$, the optimal solution of Problem~\ref{prob:stoap} is given by: 
\begin{align}
&\widehat T_{m,n}^{(t)}=
\left[\frac{1}{\theta_{m,m}}\sqrt{\frac{\rho^{(t)}\xi^{(t)}_n\left(\sum\limits_{l\neq m,l\in\mathcal{M}}\theta_{l,m}T_{l,n}^{(t-1)}+\eta_{m}\right)}{ \nu_m^{*({t})} + \rho^{(t)}\sum\limits_{\substack{j\neq m\\j\in\mathcal{M}}}\frac{\xi^{(t)}_n\theta_{m,j}T_{j,n}^{(t-1)}}{\left({\sum\limits_{l\in\mathcal{M}}{\theta_{l,j}T_{l,n}^{(t-1)}}+\eta_{j}}\right)^2}-(1-\rho^{(t)})f^{(t-1)}_{m,n} }} - \frac{\sum\limits_{\substack{l\neq m\\l\in\mathcal{M}}}\theta_{l,m}T_{l,n}^{(t-1)}+\eta_{m}}{\theta_{m,m}}\right]^1_0, \label{eqn:closed-formsto}
\end{align}
where $[x]^1_0\triangleq\min\left\{\max\{x,0\},1\right\}$ and $\nu_m^{*({t})}$ is the Lagrange multiplier that satisfies $\sum_{n\in\mathcal{N}}\widehat T_{m,n}^{(t)}=K_m$. 
\end{Lem}

\blue{Note that $\nu_m^{*({t})}$ in Lemma~\ref{lem:closed-formsto} can be efficiently obtained by using bisection search which achieves a desired accuracy $\epsilon$ with computational complexity $\mathcal O\left(M^2N\log(1/\epsilon)\right)$.}
Then, we update the caching probabilities of tier $m$ by\blue{:}
\begin{align}
&\mathbf{T}_m^{(t)}=(1-\omega^{(t)})\mathbf{T}_m^{(t-1)}+\omega^{(t)}\widehat{\mathbf{T}}_m^{(t)},\label{eqn:recTk}
\end{align}
where $\omega^{(t)}$ is a positive diminishing stepsize satisfying\blue{:}
\begin{align}
\omega^{(t)}=0,\quad \lim_{t\to\infty}\omega^{(t)}=0,\quad \sum_{t=1}^\infty\omega^{(t)}=\infty,\quad \sum_{t=1}^\infty\left(\omega^{(t)}\right)^2<\infty,\quad \lim_{t\to\infty}\frac{\omega^{(t)}}{\rho^{(t)}}=0. \label{eqn:omega}
\end{align}
Finally, the details of the proposed stochastic parallel iterative algorithm are summarized in Algorithm~\ref{alg:sto}.\footnote{\blue{Although we assume i.i.d. file request, Algorithm~\ref{alg:sto} can be applied in the case of time-variant statistics without convergence guarantee.}} Based on~\cite[Theorem~1]{7412752}, we can show the following result.

\begin{Thm}[Convergence of Algorithm~\ref{alg:sto}]\label{thm:conv-sto}
If the stepsizes $\{\rho^{(t)}\}$ and $\{\omega^{(t)}\}$ satisfy \eqref{eqn:rho} and \eqref{eqn:omega}, respectively, then every limit point of $\{\mathbf{T}^{(t)}\}$ generated by Algorithm~\ref{alg:sto} is almost surely a stationary point of Problem~\ref{prob:sto}.

\end{Thm}
\begin{IEEEproof}
Please refer to Appendix D.
\end{IEEEproof}

\begin{algorithm}[t]
    \caption{Obtaining A Stationary Point of Problem~\ref{prob:sto}}
\begin{small}
        \begin{algorithmic}[1]\label{alg:sto}
           \STATE \textbf{initialization}: choose any feasible solution $\mathbf{T}^{(0)}$ of Problem~\ref{prob:sto} as the initial point, and set $t=1$.\\
           \STATE \textbf{repeat}
           \STATE \quad observing $\{\pi_s(t):s\in\mathcal U\}$, and for all $m\in\mathcal{M}$, compute $\widehat{\mathbf{T}}_m^{(t)}$ according to \eqref{eqn:closed-formsto}, and \\ \quad update $\mathbf{T}_m^{(t)}$ according to \eqref{eqn:recTk}.
           \STATE\quad set $t=t+1$.
           \STATE \textbf{until} some convergence criteria is met.
    \end{algorithmic}
    \end{small}
\end{algorithm}

\section{Numerical Results}\label{Sec:simu}

In this section, we show the convergence of the proposed algorithms and compare the proposed algorithms with baseline schemes.
We choose $M=3$, $N = 500$,\footnote{
\blue{Although we choose file number $N=500$ for ease of simulation, numerical results show that the computational time of each proposed algorithm increases approximately linearly with $N$, demonstrating its applicability in practical networks.}}
$\alpha = 3$, \blue{$\tau_1 = 1$, $\tau_2 = 10^{-0.6}$,} \blue{$\tau_3 = 10^{-1.4}$,} $\lambda_1=3.2\times10^{-7}$, $\lambda_2=8\times10^{-6}$, $\lambda_3=8 \times10^{-4}$, $P_1=10^3P_3$ and $P_2=10^{1.4}P_3$. 
For the case of perfect file popularity distribution, we assume the file popularity follows Zipf distribution, i.e., $a_n = \frac{n^{-\gamma}}{\sum_{n\in\mathcal N}n^{-\gamma}}$, $n\in\mathcal N$, where $\gamma$ is the Zipf exponent. 
For the case of imperfect file popularity distribution, we assume the estimated file popularity is a Zipf distribution, i.e., $\widehat a_n = \frac{n^{-\widehat\gamma}}{\sum_{n\in\mathcal N}n^{-\widehat\gamma}}$, $n\in\mathcal N$, where $\widehat\gamma$ is the Zipf exponent, and \blue{choose} $\varepsilon_n=\epsilon a_n$, $n\in\mathcal N$ for some $\epsilon\in(0,1)$. 
For the case of unknown file popularity distribution, we assume that file requests follow the Zipf distribution, i.e., $a_n = \frac{n^{-\gamma}}{\sum_{n\in\mathcal N}n^{-\gamma}}$, $n\in\mathcal N$, as in the case of perfect file popularity. 
\blue{In addition, we choose ${\rm Pr}[\pi_s(t)\neq0]=0.9$. Note that in the simulation, we focus on serving only cached files. When $K_m/N$, $m\in\mathcal M$ are small, file availability in the $M$-tier network is low, leading to small STPs.}

\subsection{Convergence}
\begin{figure}[t]
\begin{center}
{\resizebox{9cm}{!}{\includegraphics{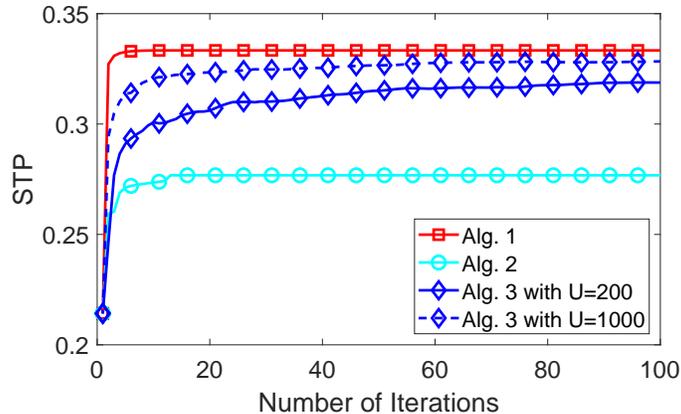}}}
\end{center}
\vspace{-3mm}
\caption{\small{STP, Worst-case STP and stochastic STP versus number of iterations in the three cases with $K_1=80$, $K_2=60$, $K_3=40$, $\gamma=\widehat\gamma=0.55$ and $\epsilon=0.25$. We choose the same initial point for all the algorithms.}}
\vspace{-3mm}
\label{fig:simulation-convergence}
\end{figure}

First, we show the convergence of the proposed algorithms. Fig.~\ref{fig:simulation-convergence} illustrates the STP, worst-case STP and stochastic STP versus the number of iterations. 
\blue{Although the number of iterations of our proposed algorithms cannot be analytically characterized, from Fig.~\ref{fig:simulation-convergence}, we can observe that our proposed algorithms terminate in a few iterations.
Note that different form Algorithm~1 and Algorithm~2, which are based on the perfect and imperfect popularity distributions at the initial stage, Algorithm~3 can make use of instantaneous file requests observed at each slot and gradually improve the random caching design. In addition, note that Problem 1 and Problem 3 share the same sets of optimal solutions and stationary points, as $\mathbb{E}\left[q(\boldsymbol\xi,\mathbf T)\right]=q(\mathbf{a},\mathbf{T})$. Hence, the converged STPs of Algorithm~1 and Algorithm~3 are similar and are larger than the STP of Algorithm.~2.}
In addition, Algorithm~\ref{alg:sto} with a larger $U$ converges faster, as \blue{at each slot} more observations for file requests can be used to approximate the file popularity distribution and the gradient of the stochastic STP.

\subsection{Performance}
Next, we compare the STP of Algorithm~\ref{alg:rand} (obtained using 20 iterations), the STP of Algorithm~\ref{alg:ro} (obtained using 30 iterations) and the STP of Algorithm~\ref{alg:sto} with \blue{$U=200$} (obtained using \blue{200} iterations (slots)) with those of three baseline schemes.
Baseline~1 (most popular) refers to the design in which each BS in tier $m$ stores the $K_m$ most popular files~\cite{bacstuug2015transfer} according to ${\mathbf a}$ in the case of perfect file popularity distribution, $\widehat{\mathbf a}$ in the case of imperfect file popularity distribution, and the estimated file popularity $\widetilde{\mathbf a}\triangleq(\widetilde a_n)_{n\in\mathcal N}$ in the case of unknown file popularity distribution, where \blue{$\widetilde a_n\triangleq\frac{\sum_{t=1}^{L}\sum_{s\in\mathcal U}\mathbf I\left[\pi^{(t)}_s=n\right]}{\sum_{t=1}^{L}\sum_{s\in\mathcal U}\mathbf I\left[\pi^{(t)}_s\neq0\right]}$, $n\in\mathcal N$, are obtained based on the accumulated file requests within $L$ slots.}
Baseline~2 (i.i.d. file popularity) refers to the design in which each BS in tier $m$ randomly stores $K_m$ files, in an i.i.d. manner~\cite{Bharath} with file $n$ being selected with probabilities ${a}_n$ in the case of perfect file popularity distribution, $\widehat{a}_n$ in the case of imperfect file popularity distribution, and $\widetilde{a}_n$ in the case of unknown file popularity distribution. 
\blue{Baseline~3 (suboptimal) refers to the suboptimal caching solution proposed in \cite{wen2016cache} based on $\mathbf{a}_n$ in the case of perfect file popularity distribution, $\widehat{\mathbf{a}}_n$ in the case of imperfect file popularity distribution, and $\widetilde{\mathbf{a}}_n$ in the case of unknown file popularity distribution.}
\blue{In addition, we compare with Algorithm~1 based on $\hat{\mathbf{a}}$ in the case of imperfect file popularity distribution, referred to as Alg.~1 (imperfect), and with Algorithm~1 based on $\widetilde{\mathbf{a}}$ in the case of unknown file popularity, referred to as Alg.~1 (stochastic).}\footnote{
\blue{Please note that although Baseline 1 (most popular) and Baseline 2 (i.i.d. file popularity) are not optimization-based, they are widely used as baseline schemes in existing works on wireless caching \cite{ICC15Giovanidis,cui2016analysis,cui2017analysis,li2016optimization,wen2016cache,Wang2017Joint}. In addition, note that the proposed solutions in~\cite{li2016optimization} and~\cite{Wang2017Joint} are for the case of uniform SIR threshold and for a two-tier network, respectively, and hence cannot be applied in our setup.}}

\begin{figure}[t]
\begin{center}
\subfigure[\scriptsize{
cache size $K_3$ at $\gamma=0.55$, $K_1=K_3+40$, $K_2=K_3+20$.}\label{fig:simulation-pe-size}]
{\resizebox{6cm}{!}{\includegraphics{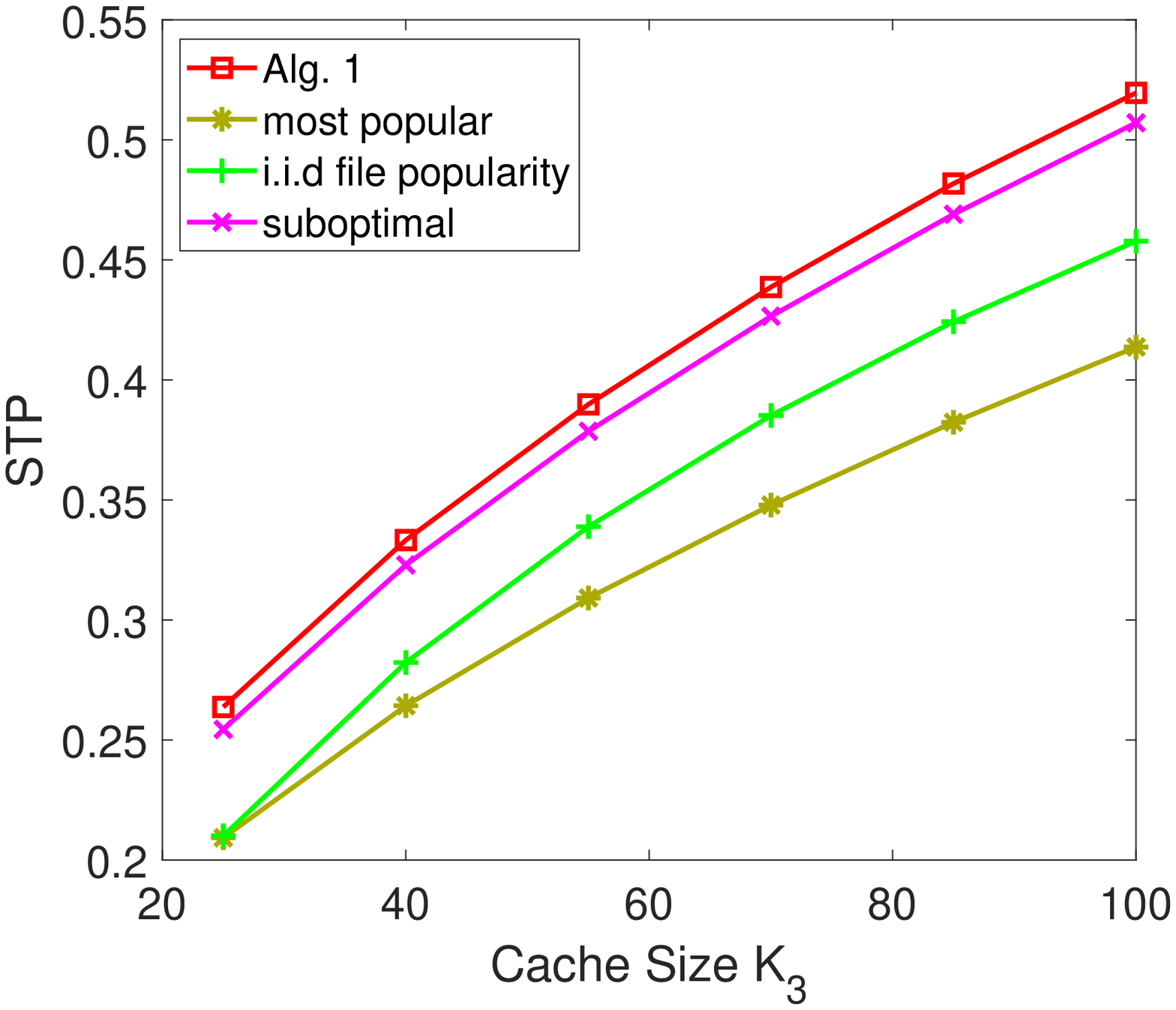}}}\quad
\subfigure[\scriptsize{
Zipf exponent $\gamma$ at $K_1=80$, $K_2=60$, $K_3=40$.}\label{fig:simulation-pe-zipf}]
{\resizebox{6cm}{!}{\includegraphics{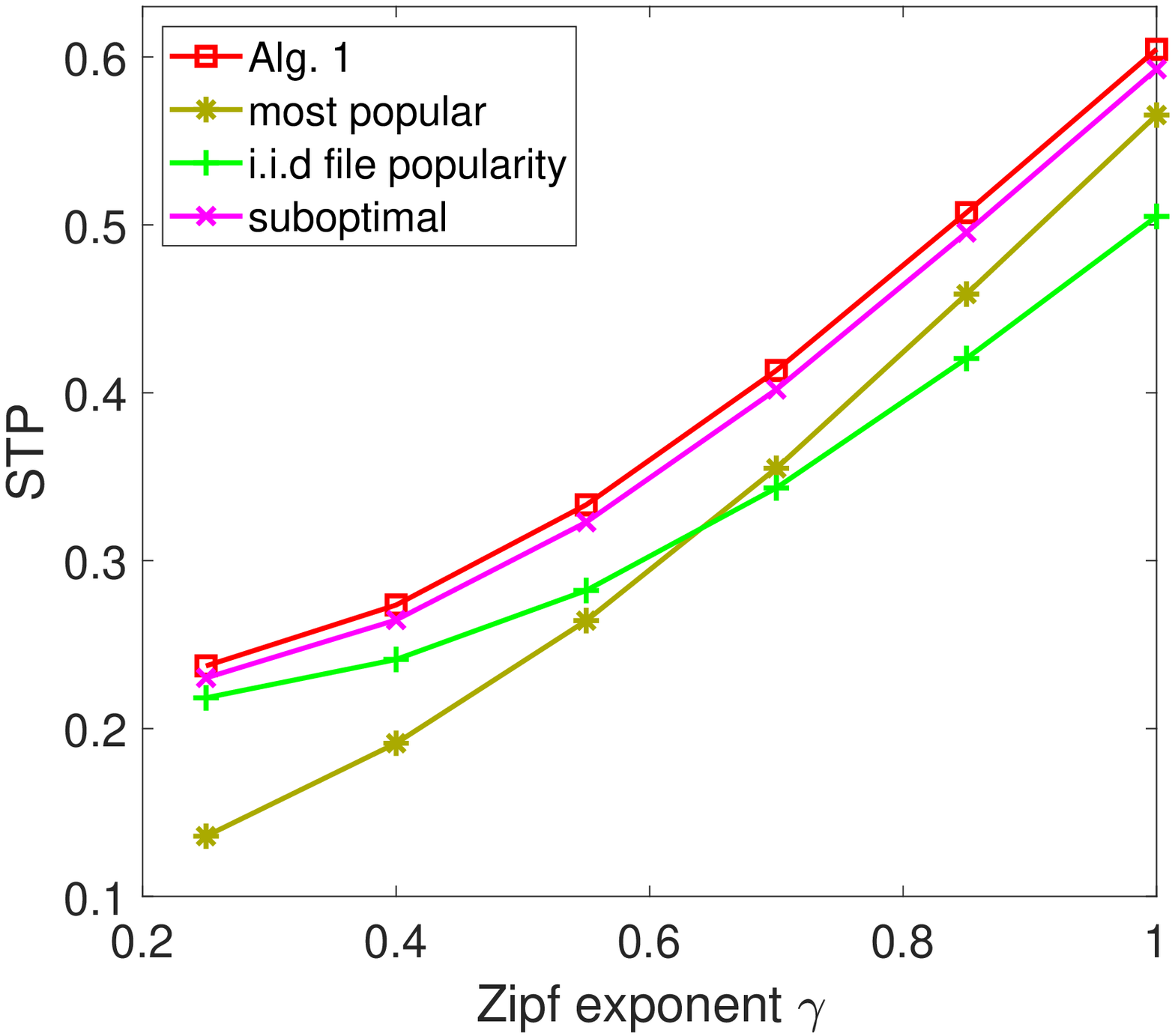}}}
\end{center}
\vspace{-2mm}
\caption{\small{STP versus cache sizes and Zipf exponent $\gamma$ in the case of perfect file popularity distribution.}}
\vspace{-2mm}
\label{fig:simulation-performance-pe}
\end{figure}

\begin{figure}[t]
\begin{center}
\subfigure[\scriptsize{
cache size $K_3$ at $\widehat\gamma=0.55$, $\epsilon=0.25$, $K_1=K_3+40$, $K_2=K_3+20$.}\label{fig:simulation-ro-size}]
{\resizebox{6cm}{!}{\includegraphics{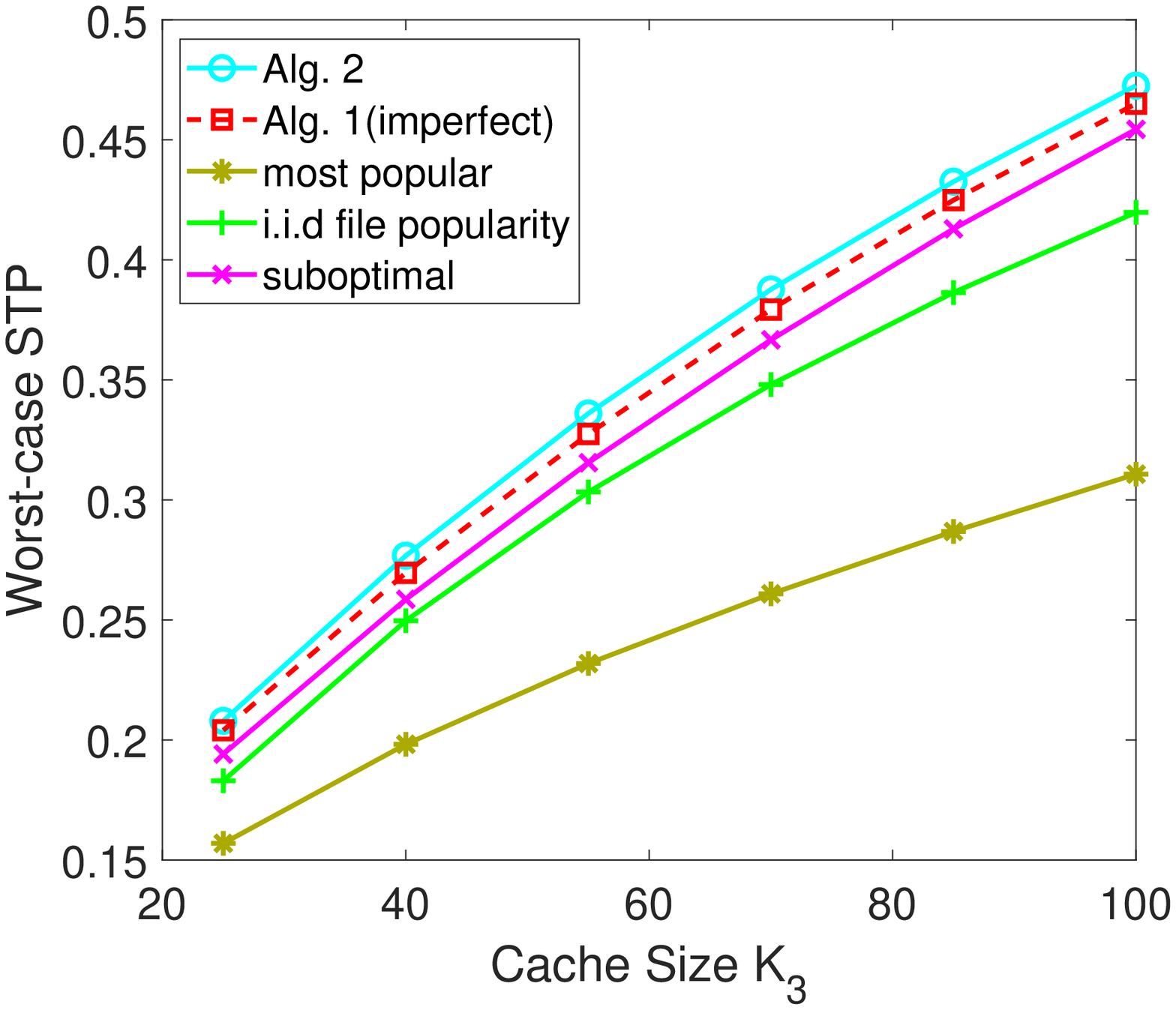}}}\quad
\subfigure[\scriptsize{
Zipf exponent $\widehat\gamma$ at $\epsilon=0.25$, $K_1=80$, $K_2=60$, $K_3=40$.}\label{fig:simulation-ro-zipf}]
{\resizebox{6cm}{!}{\includegraphics{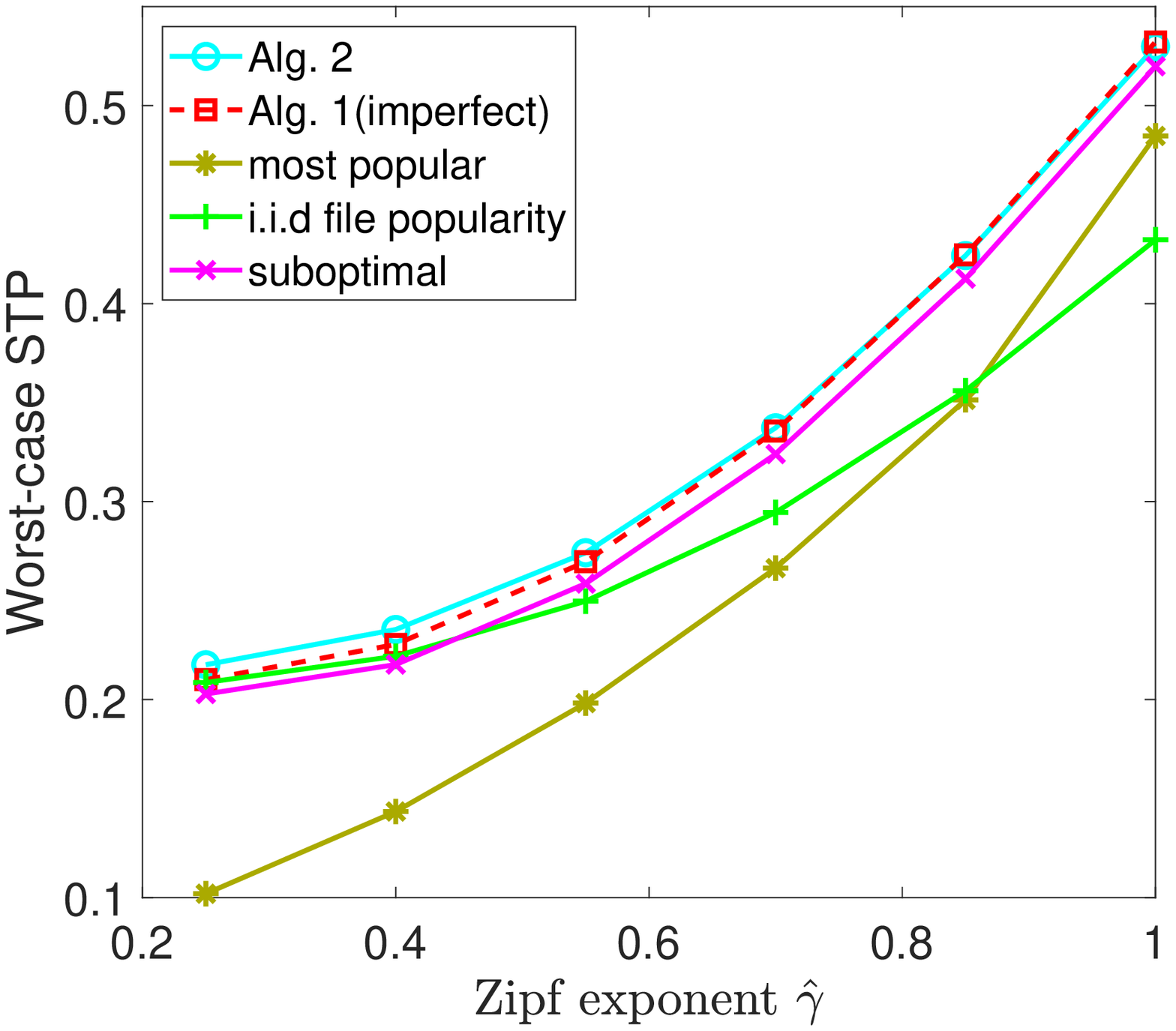}}}\quad
\subfigure[\scriptsize{
Error bound parameter $\epsilon$ at $\widehat\gamma=0.55$, $K_1=80$, $K_2=60$, $K_3=40$.}\label{fig:simulation-ro-error}]
{\resizebox{6cm}{!}{\includegraphics{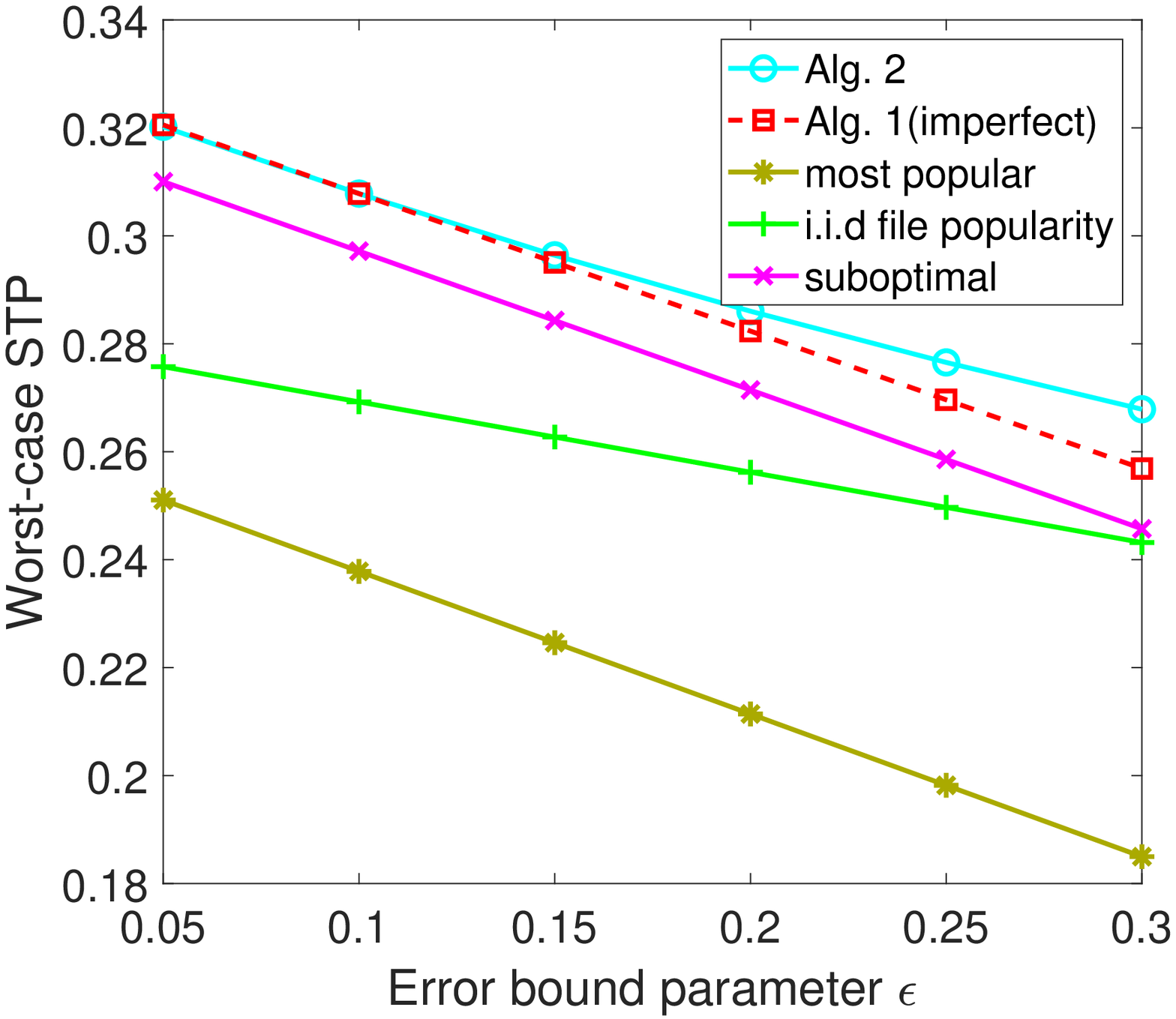}}}
\end{center}
\vspace{-2mm}
\caption{\small{Worst-case STP versus cache sizes, Zipf exponent $\widehat\gamma$, and error bound parameter $\epsilon$ in the case of imperfect file popularity distribution.}}
\vspace{-2mm}
\label{fig:simulation-performance-ro}
\end{figure}

\begin{figure}[t]
\begin{center}
\subfigure[\scriptsize{
cache size $K_3$ at $\gamma=0.55$, $L=30$, $K_1=K_3+40$, $K_2=K_3+20$.}\label{fig:simulation-sto-size}]
{\resizebox{6cm}{!}{\includegraphics{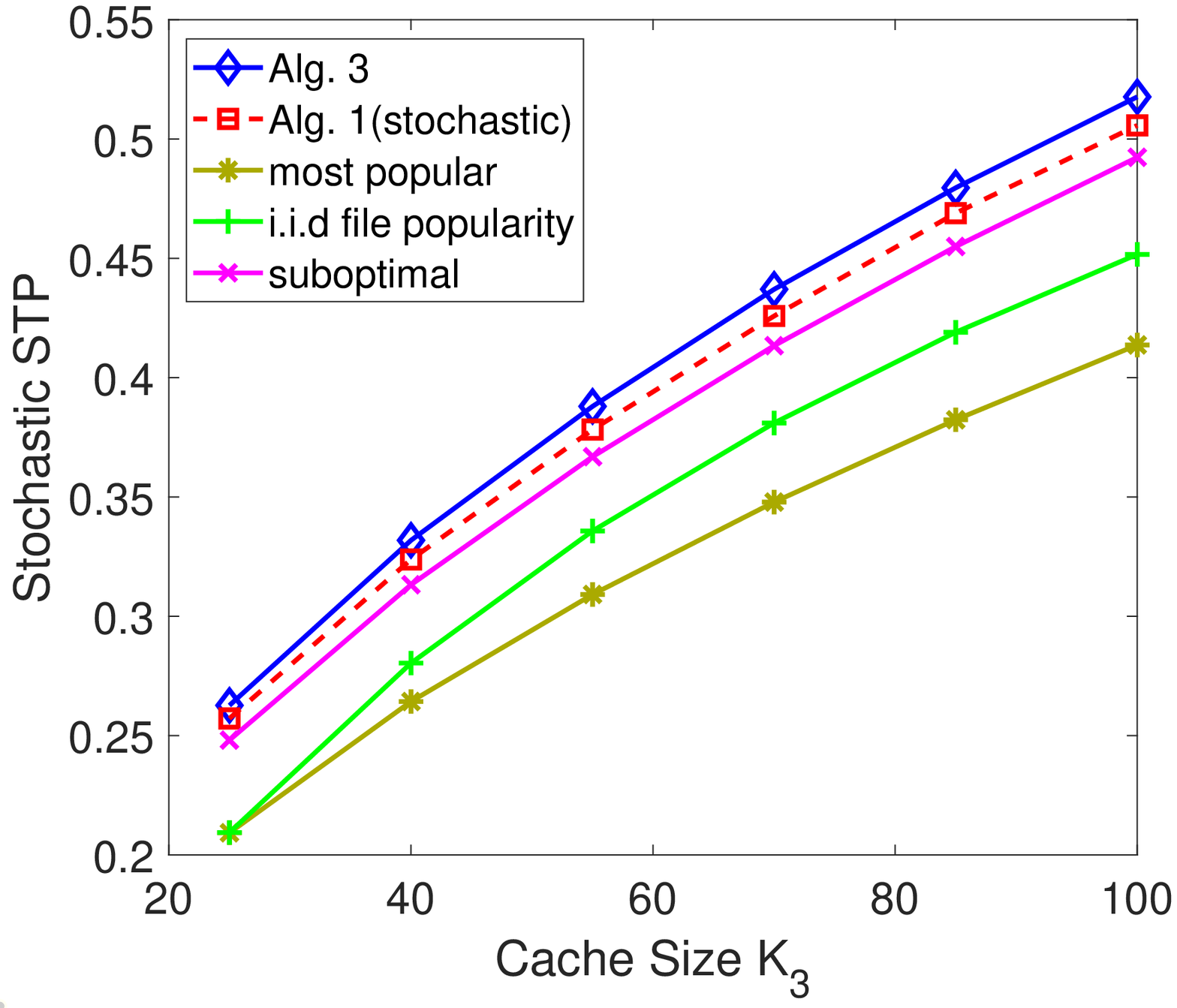}}}\quad
\subfigure[\scriptsize{
Zipf exponent $\gamma$ at $L=30$, $K_1=80$, $K_2=60$, $K_3=40$.}\label{fig:simulation-sto-zipf}]
{\resizebox{6cm}{!}{\includegraphics{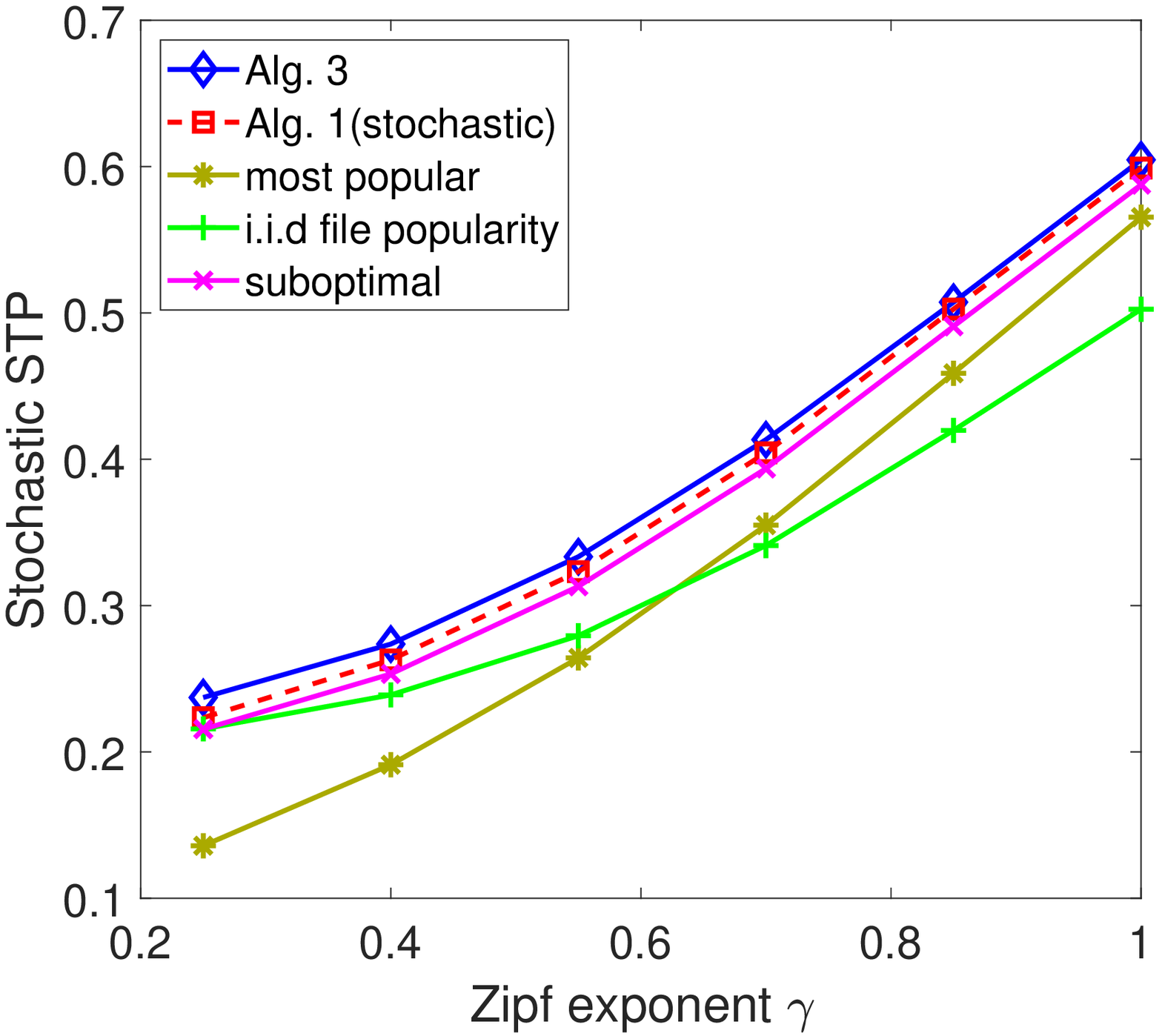}}}
\subfigure[\scriptsize{
Number of slots $L$ at $\gamma=0.55$, $K_1=80$, $K_2=60$, $K_3=40$.}\label{fig:simulation-sto-slot}]
{\resizebox{6cm}{!}{\includegraphics{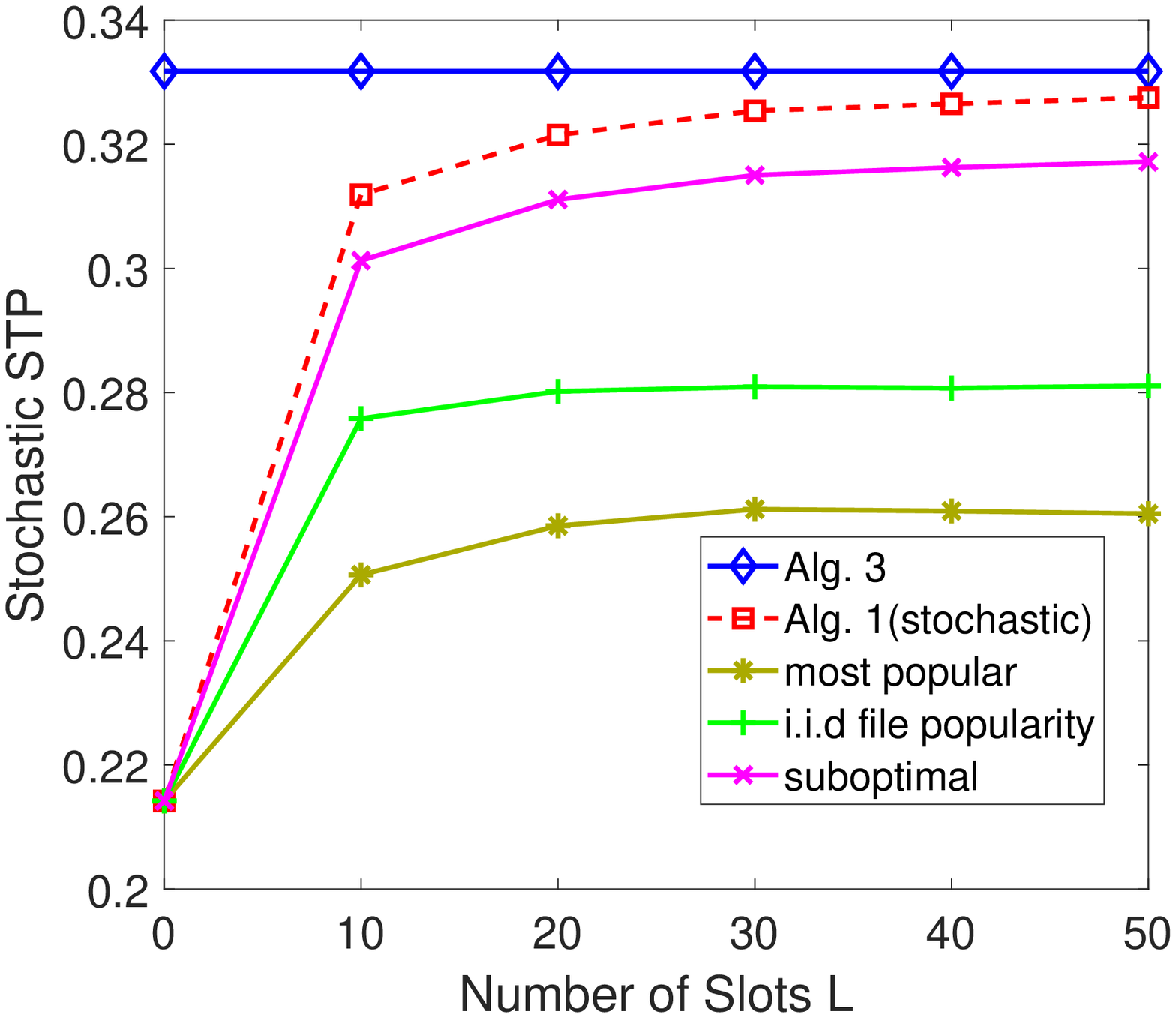}}}
\end{center}
\vspace{-2mm}
\caption{\small{Stochastic STP versus cache sizes, Zipf exponent $\gamma$ and number of time slots $L$ in the case of unknown file popularity distribution.}}
\vspace{-2mm}
\label{fig:simulation-performance-sto}
\end{figure}

Fig.~\ref{fig:simulation-pe-size}, Fig.~\ref{fig:simulation-ro-size} and Fig.~\ref{fig:simulation-sto-size} illustrate the STP, the worst-case STP and the stochastic STP versus the cache sizes, respectively. From Fig.~\ref{fig:simulation-pe-size}, Fig.~\ref{fig:simulation-ro-size} and Fig.~\ref{fig:simulation-sto-size}, we can observe that as the cache sizes $K_m$, $m\in\mathcal M$ increase, the STP of each scheme increases. This is because as $K_m$, $m\in\mathcal M$ increase, each BS can cache more files, and the probability that a randomly requested file is cached at a nearby BS increases. In addition, note that for each scheme, its STP in Fig.~\ref{fig:simulation-pe-size} is very close to its stochastic STP in Fig.~\ref{fig:simulation-sto-size}, since under the same file popularity distribution, its stochastic STP converges to its STP as the number of \blue{iterations} goes to infinity.

Fig.~\ref{fig:simulation-pe-zipf}, Fig.~\ref{fig:simulation-ro-zipf} and Fig.~\ref{fig:simulation-sto-zipf} illustrate the STP, the worst-case STP and the stochastic STP versus the Zipf exponents, respectively. From Fig.~\ref{fig:simulation-pe-zipf}, Fig.~\ref{fig:simulation-ro-zipf} and Fig.~\ref{fig:simulation-sto-zipf}, we can observe that as the Zipf exponent $\gamma$ $(\widehat\gamma)$ increases, the STP of each scheme increases. This is because when $\widehat\gamma$ ($\gamma$) increases, the tail of the Zipf distribution becomes small, and hence, the probability that a randomly requested file is cached at a nearby BS increases for each scheme. Similarly, note that for each scheme, its STP in Fig.~\ref{fig:simulation-pe-zipf} is very close to its stochastic STP in Fig.~\ref{fig:simulation-sto-zipf}.

Fig.~\ref{fig:simulation-ro-error} illustrates the worst-case STP versus the error bound parameter $\epsilon$ in the case of imperfect file popularity. From Fig.~\ref{fig:simulation-ro-error}, we can see that as $\epsilon$ increases, the worst-case STP of each scheme decreases. The reason is that as $\epsilon$ increases, the possible estimation error increases, and the worst-case STP decreases. In addition, we can observe that the decrease rates of the worst-case STPs of all schemes are different, as the estimation error has different impacts on these schemes. 
\blue{Finally, the STP gaps between Algorithm 2 and the other schemes increase with the error bound parameter, as those schemes do not consider the uncertainty in the file popularity estimation. This reveals the value of robust caching optimization in the case of imperfect file popularity distribution.}

\blue{Fig.~\ref{fig:simulation-sto-slot} illustrates the stochastic STP versus the number of slots $L$ which is used for estimating $\tilde{\mathbf{a}}$ in the case of unknown file popularity. 
From Fig.~\ref{fig:simulation-sto-slot}, we can see that as $L$ increases, the stochastic STPs of all schemes except for Algorithm 3 increase. The reason is that as $L$ increases, the estimation of the file popularity distribution becomes more accurate. In addition, we can observe that the stochastic STP of Algorithm 3 is not adaptive to $L$, as the proposed solution can be continuously improved by accumulating new file requests. 
Finally, the STP gaps between Algorithm 3 and the other schemes increase as the number of slots $L$ used for estimating $\tilde{\mathbf{a}}$ decreases, 
revealing the value of stochastic caching optimization in the case of unknown file popularity distribution.
}

Finally, from Fig~\ref{fig:simulation-performance-pe}, Fig~\ref{fig:simulation-performance-ro} and Fig~\ref{fig:simulation-performance-sto}, we can observe that our proposed solutions outperform the baseline schemes in all three cases.
This is because in each case, the properties of the file popularity distribution are appropriately captured when formulating the optimal caching design problem, and a stationary point of the challenging optimization problem is obtained using the proposed algorithm. \blue{Note that each point corresponding to the proposed solution in Fig~\ref{fig:simulation-performance-pe}, Fig~\ref{fig:simulation-performance-ro} and Fig~\ref{fig:simulation-performance-sto} is obtained with an arbitrary initial point, showing that the performance of a stationary point is promising at least under our setup.}

\section{Conclusion}

In this paper, we considered optimal \blue{random} caching designs for perfect, imperfect and unknown file popularity distributions in  large-scale multi-tier wireless networks.
First, in the case of perfect file popularity distribution, we formulate \blue{the successful transmission probability (STP) optimization problem}, which is nonconvex. We \blue{propose} an efficient parallel iterative algorithm to obtain a stationary point.
Then, in the case of imperfect file popularity distribution, we formulate \blue{the worst-case STP maximization problem. This} is a challenging robust optimization problem, \blue{and} we \blue{propose} an efficient iterative algorithm \blue{to obtain} a stationary point.
Next, in the case of unknown file popularity distribution, we formulate \blue{the stochastic STP (i.e., the STP in the stochastic form) maximization problem. This} is a challenging nonconvex stochastic optimization problem, \blue{and} we develop an efficient iterative algorithm to obtain a stationary point.
Finally, by numerical results, we show that the proposed solutions achieve \blue{notable} gains over existing schemes in all three cases. 

\section*{Appendix A: Proof of Theorem~\ref{thm:conv-rand}}
We show that the assumptions in~\cite[Theorem~1]{razaviyayn2014parallel} are satisfied. 
\begin{itemize}
\item It is clear that $h_m\big(\mathbf{a},\mathbf{T}_m,\mathbf{T}^{(k)}\big)$ is continuously differentiable for any given $\mathbf{a}$ and $\mathbf{T}^{(k)}$. 
It remains to show that $h_m\big(\mathbf{a},\mathbf{T}_m,\mathbf{T}^{(k)}\big)$ is strongly convex in $\mathbf T_m$. Note that the Hessian of $h_m\big(\mathbf{a},\mathbf{T}_m,\mathbf{T}^{(k)}\big)$ is $\boldsymbol\nabla^2_{\mathbf T_m}h_m\big(\mathbf{a},\mathbf{T}_m,\mathbf{T}^{(k)}\big)\triangleq\Big(\frac{\partial^2 h_m\big(\mathbf{a},\mathbf{T}_m,\mathbf{T}^{(k)}\big)}{\partial T_{m,n}\partial T_{m,n'}}\Big)_{n\in\mathcal N,n'\in\mathcal N}$, where
\begin{align}
\frac{\partial^2 h_m\big(\mathbf{a},\mathbf{T}_m,\mathbf{T}^{(k)}\big)}{\partial T_{m,n}\partial T_{m,n'}}=
\begin{cases}
{\frac{-2 a_n \eta_m \theta_{m,m}}{\left({\theta_{m,m}T_{m,n}}+\sum_{l\in\mathcal{M},l\neq m}{\theta_{l,m}T^{(k)}_{l,n}}+\eta_{m}\right)^3}},\quad &n'= n,\\
0,\quad &n'\neq n.
\end{cases}
\end{align} 
Thus, $\boldsymbol\nabla^2_{\mathbf T_m}h_m\big(\mathbf{a},\mathbf{T}_m,\mathbf{T}^{(k)}\big)\preceq-b\mathbf E$, where $b\triangleq\min\limits_{n\in\mathcal N}{\frac{2 a_n \eta_m \theta_{m,m}}{\left({\theta_{m,m}}+\sum_{l\in\mathcal{M},l\neq m}{\theta_{l,m}T^{(k)}_{l,n}}+\eta_{m}\right)^3}}$, and $\mathbf E$ is the identity matrix. That is, $h_m\big(\mathbf{a},\mathbf{T}_m,\mathbf{T}^{(k)}\big)$ is uniformly strongly concave with constant $b>0$. Thus, the first assumption of~\cite[Theorem~1]{razaviyayn2014parallel} is satisfied.
\item By \eqref{eqn:randap}, we have $\nabla_{\mathbf{T}_{m}}{h_m\big(\mathbf{a},\mathbf{T}_m,\mathbf{T}^{(k)}\big)}=\nabla_{\mathbf{T}_{m}}{q\big(\mathbf a,\mathbf T\big)}$ for all $m\in\mathcal M$. Thus, the second assumption of~\cite[Theorem~1]{razaviyayn2014parallel} is satisfied.
\item It is clear that $h_m\big(\mathbf{a},\mathbf{T}_m,\mathbf{T}^{(k)}\big)$ is smooth on the constraint set determined by \eqref{eqn:T1} and \eqref{eqn:T2} for any given $\mathbf{a}$ and $\mathbf{T}_m$, and hence its derivative is Lipschitz continuous. Thus, the third assumption of~\cite[Theorem~1]{razaviyayn2014parallel} is satisfied.
\end{itemize}
Therefore, Theorem~\ref{thm:conv-sto} readily follows from~\cite[Theorem~1]{razaviyayn2014parallel}.

\section*{Appendix B: Proof of Lemma~\ref{lem:roeq}}
Firstly, the inner problem of Problem~\ref{prob:ro}, $\min_{\mathbf{a}\in\mathcal{A}}q_{\infty}\left(\mathbf a,\mathbf{T}\right)$, can be rewritten as follows.
\begin{Prob}[Inner LP of Problem~\ref{prob:ro}]\label{prob:inter}
	\begin{align}
	 \min_{\mathbf a} \quad &q(\mathbf a,\mathbf T)\nonumber\\
\text{s.t.} \quad
&\underline a_n\leq a_n,\quad n\in\mathcal{N},\label{eqn:alowbound}\\
&a_n\leq\overline a_n,\quad n\in\mathcal{N},\label{eqn:aupbound}\\
&\sum_{n\in\mathcal N}a_n=1. \label{eqn:asum}
	\end{align}
\end{Prob}

It is clear that Problem~\ref{prob:inter} is an LP with respect to $\mathbf a$ for any given $\mathbf T$, satisfying \eqref{eqn:T1} and \eqref{eqn:T2}. From \cite[pp.225]{boyd2004convex}, we obtain its dual problem as below.
\begin{Prob}[Dual Problem of Inner LP of Problem~\ref{prob:ro}]\label{prob:interdual}
	\begin{align}
	 \max_{\boldsymbol{\lambda},\boldsymbol{\mu},{\nu}} \quad &\sum_{n\in\mathcal{N}}{(\lambda_n \underline a_n-\mu_n \overline a_n)}-\nu\nonumber\\
\text{s.t.} \quad &\eqref{eqn:dual1},\nonumber\\
&\lambda_n\geq 0, \quad n\in\mathcal{N},\label{eqn:dual20}\\
&\mu_n\geq 0, \quad n\in\mathcal{N},\label{eqn:dual30}
	\end{align}
\end{Prob}
where $\boldsymbol{\lambda}$, $\boldsymbol{\mu}$ and $\nu$ are the Lagrange multipliers corresponding to \eqref{eqn:alowbound}, \eqref{eqn:aupbound} and \eqref{eqn:asum}, respectively.

As strong duality holds for LP, the dual problem in Problem~\ref{prob:interdual} and the primal problem in Problem~\ref{prob:inter} share the same optimal value, which can be viewed as a function of $\mathbf T$. Thus, the maximin problem in Problem~\ref{prob:ro} can be equivalently converted to the maximization problem in Problem~\ref{prob:dualmax}, by replacing the inner LP problem in Problem~\ref{prob:inter} with its dual problem in Problem~\ref{prob:interdual}.

\section*{Appendix C: Proof of Theorem~\ref{thm:conv-ro}}
By~\cite[Proposition~3]{4275017}, it can be easily shown that Algorithm~\ref{alg:ro} converges to a stationary point of Problem~\ref{prob:roep}. 
Let $\left(\mathbf{T}^{\star}\right.$, $\mathbf{x}^{\star}$, $\boldsymbol{\lambda}^{\star}$, $\boldsymbol{\mu}^{\star}$, ${\nu_1}^{\star}$, $\left.{\nu_2}^{\star}, y^{\star}\right)$ denote a stationary point of Problem~\ref{prob:roep}, which satisfies the KKT conditions of Problem~\ref{prob:roep}. It remains to show that $\left(\mathbf{T}^{\star}\right.$, $\boldsymbol{\lambda}^{\star}$, $\boldsymbol{\mu}^{\star}$, $\left.{\nu_1}^{\star}-{\nu_2}^{\star}\right)$ is a stationary point of Problem~\ref{prob:dualmax}, which satisfies the KKT conditions of Problem~\ref{prob:dualmax}.

First, we derive the KKT conditions of Problem~\ref{prob:roep} based on its equivalent form given as follows.
\begin{Prob}[Equivalent Form of Problem~\ref{prob:roep}]\label{prob:roep2}
\begin{align}
\min_{{\mathbf{T},\mathbf{x},\boldsymbol{\lambda},\boldsymbol{\mu}, {\nu_1},{\nu_2},y}} \quad &-y \nonumber\\
\text{s.t.} \quad &\eqref{eqn:dual20},\eqref{eqn:dual30},\nonumber\\
&y-\sum_{n\in\mathcal{N}}{(\lambda_n \underline a_n-\mu_n \overline a_n)}+(\nu_1-\nu_2)\leq0\label{eqn:provprob1}\\
&\lambda_n-\mu_n-(\nu_1-\nu_2)-\sum_{m\in\mathcal{M}} {T_{m,n}}{x_{m,n}^{-1}}\leq0, \quad n\in\mathcal{N},\label{eqn:provprob2}\\
&\left({\sum_{l\in\mathcal{M}} \theta_{l,m} T_{l,n} + \eta_{m}}\right){x^{-1}_{m,n}}\leq1, \quad m\in\mathcal{M},\ n\in\mathcal{N}, \label{eqn:provprob3}\\
&T_{m,n}\geq 0,\quad m\in\mathcal M,\ n\in\mathcal N, \label{eqn:provprob4}\\
&T_{m,n}\leq 1,\quad m\in\mathcal M,\ n\in\mathcal N, \label{eqn:provprob5}\\
&\sum\limits_{n\in\mathcal N}T_{m,n}\leq K_m,\quad m\in\mathcal M. \label{eqn:provprob6}
\end{align}
\end{Prob}
The Lagrangian function of Problem~\ref{prob:roep2} is given by:
\begin{align}
&\mathcal{L}\left(\mathbf{T},\mathbf{x}, \boldsymbol{\lambda}, \boldsymbol{\mu}, {\nu_1}, {\nu_2}, y, \widetilde{\boldsymbol\nu}_1, \widetilde{\boldsymbol\nu}_2, \widetilde\lambda_0, \widetilde{\boldsymbol\lambda}, \widetilde{\boldsymbol\mu}, \widetilde{\boldsymbol\kappa}_1, \widetilde{\boldsymbol\kappa}_2, \widetilde{\boldsymbol\iota}\right)=
-y+\sum_{n\in\mathcal N}\widetilde\nu_{1,n}(-\lambda_n)+\sum_{n\in\mathcal N}\widetilde\nu_{2,n}(-\mu_n)\nonumber\\
&+\widetilde\lambda_0\left(y-\sum_{n\in\mathcal{N}}{(\lambda_n \underline a_n-\mu_n \overline a_n)}+(\nu_1-\nu_2)\right)
+\sum_{n\in\mathcal N}\widetilde\lambda_n\left(\lambda_n-\mu_n-(\nu_1-\nu_2)-\sum_{m\in\mathcal{M}} {T_{m,n}}{x_{m,n}^{-1}}\right)\nonumber\\
&+\sum_{m\in\mathcal M}\sum_{n\in\mathcal N}\widetilde\mu_{m,n}\left(\left({\sum_{l\in\mathcal{M}} \theta_{l,m} T_{l,n} + \eta_{m}}\right){x^{-1}_{m,n}}-1\right)
+\sum_{m\in\mathcal M}\widetilde\iota_m\left(\sum\limits_{n\in\mathcal N}T_{m,n}-K_m\right)\nonumber\\
&+\sum_{m\in\mathcal M}\sum_{n\in\mathcal N}\widetilde\kappa_{1,m,n}(-T_{m,n})
+\sum_{m\in\mathcal M}\sum_{n\in\mathcal N}\widetilde\kappa_{2,m,n}(T_{m,n}-1), \label{eqn:lag}
\end{align}
where $\widetilde{\boldsymbol\nu}_1\triangleq(\widetilde\nu_{1,n})_{n\in\mathcal N}$, $\widetilde{\boldsymbol\nu}_2\triangleq(\widetilde\nu_{2,n})_{n\in\mathcal N}$, $\widetilde\lambda_0$, $\widetilde{\boldsymbol\lambda}\triangleq(\widetilde\lambda_{n})_{n\in\mathcal N}$, $\widetilde{\boldsymbol\mu}\triangleq(\widetilde\mu_{m,n})_{m\in\mathcal M,n\in\mathcal N}$, $\widetilde{\boldsymbol\kappa}_1\triangleq(\widetilde\kappa_{1,m,n})_{m\in\mathcal M,n\in\mathcal N}$, $\widetilde{\boldsymbol\kappa}_2\triangleq(\widetilde\kappa_{2,m,n})_{m\in\mathcal M,n\in\mathcal N}$ and $\widetilde{\boldsymbol\iota}\triangleq(\widetilde\iota_{m})_{m\in\mathcal M}$ are the Lagrange multipliers corresponding to \eqref{eqn:dual20}, \eqref{eqn:dual30}, \eqref{eqn:provprob1}, \eqref{eqn:provprob2}, \eqref{eqn:provprob3}, \eqref{eqn:provprob4}, \eqref{eqn:provprob5} and \eqref{eqn:provprob6}, respectively.

Next, as $\left(\mathbf{T}^{\star}\right.$, $\mathbf{x}^{\star}$, $\boldsymbol{\lambda}^{\star}$, $\boldsymbol{\mu}^{\star}$, ${\nu_1}^{\star}$, ${\nu_2}^{\star}$, $\left.y^{\star}\right)$ is a stationary point of Problem~\ref{prob:roep}, there exists a dual point $\left(\widetilde{\boldsymbol\nu}_1^{\star}\right.$, $\widetilde{\boldsymbol\nu}_2^{\star}$, $\widetilde\lambda_0^{\star}$, $\widetilde{\boldsymbol\lambda}^{\star}$, $\widetilde{\boldsymbol\mu}^{\star}$, $\widetilde{\boldsymbol\kappa}_1^{\star}$, $\widetilde{\boldsymbol\kappa}_2^{\star}$, $\left.\widetilde{\boldsymbol\iota}^{\star}\right)$ together with $\left(\mathbf{T}^{\star}\right.$, $\mathbf{x}^{\star}$, $\boldsymbol{\lambda}^{\star}$, $\boldsymbol{\mu}^{\star}$, ${\nu_1}^{\star}$, ${\nu_2}^{\star}$, $\left.y^{\star}\right)$ satisfies the following KKT conditions:
\begin{itemize}
\item Feasibility: The primal point $\left(\mathbf{T}^{\star}\right.$, $\mathbf{x}^{\star}$, $\boldsymbol{\lambda}^{\star}$, $\boldsymbol{\mu}^{\star}$, ${\nu_1}^{\star}$, $\left.{\nu_2}^{\star}, y^{\star}\right)$ satisfies the primal constraints in \eqref{eqn:dual20}, \eqref{eqn:dual30}, \eqref{eqn:provprob1}, \eqref{eqn:provprob2}, \eqref{eqn:provprob3}, \eqref{eqn:provprob4}, \eqref{eqn:provprob5} and \eqref{eqn:provprob6}, and the dual point $\left(\widetilde{\boldsymbol\nu}_1^{\star}\right.$, $\widetilde{\boldsymbol\nu}_2^{\star}$, $\widetilde\lambda_0^{\star}$, $\widetilde{\boldsymbol\lambda}^{\star}$, $\widetilde{\boldsymbol\mu}^{\star}$, $\widetilde{\boldsymbol\kappa}_1^{\star}$, $\widetilde{\boldsymbol\kappa}_2^{\star}$, $\left.\widetilde{\boldsymbol\iota}^{\star}\right)$ satisfies the dual constraints.
\begin{align}
\widetilde\nu_{1,n}^{\star}, \widetilde\nu_{2,n}^{\star}, \widetilde\lambda_0^{\star}, \widetilde\lambda_n^{\star}, \widetilde\mu_{m,n}^{\star}, \widetilde\kappa_{1,m,n}^{\star}, \widetilde\kappa_{2,m,n}^{\star}, \widetilde\iota_{m}^{\star}\geq0, \quad m\in\mathcal M,\ n\in\mathcal N. \label{eqn:dual-feasible}
\end{align}

\item Complementary slackness: $\left(\mathbf{T}^{\star}\right.$, $\mathbf{x}^{\star}$, $\boldsymbol{\lambda}^{\star}$, $\boldsymbol{\mu}^{\star}$, ${\nu_1}^{\star}$, $\left.{\nu_2}^{\star}, y^{\star}\right)$ and $\left(\widetilde{\boldsymbol\nu}_1^{\star}\right.$, $\widetilde{\boldsymbol\nu}_2^{\star}$, $\widetilde\lambda_0^{\star}$, $\widetilde{\boldsymbol\lambda}^{\star}$, $\widetilde{\boldsymbol\mu}^{\star}$, $\widetilde{\boldsymbol\kappa}_1^{\star}$, $\widetilde{\boldsymbol\kappa}_2^{\star}$, $\left.\widetilde{\boldsymbol\iota}^{\star}\right)$ satisfy:
\begin{align}
&\widetilde\nu_{1,n}^{\star}\lambda_n^{\star}=0,\quad {n\in\mathcal N},\label{eqn:lambdacs}\\
&\widetilde\nu_{2,n}^{\star}\mu_n^{\star}=0,\quad {n\in\mathcal N},\label{eqn:mucs}\\
&\widetilde\lambda_0^{\star}\left(y^{\star}-\sum_{n\in\mathcal{N}}{(\lambda_n^{\star} \underline a_n-\mu_n^{\star} \overline a_n)}+(\nu_1^{\star}-\nu_2^{\star})\right)=0,\\
&\widetilde\lambda_n^{\star}\left(\lambda_n^{\star}-\mu_n^{\star}-(\nu_1^{\star}-\nu_2^{\star})-\sum_{m\in\mathcal{M}} {T_{m,n}^{\star}}({x_{m,n}^{\star}})^{-1}\right)=0,\quad {n\in\mathcal N},\label{eqn:slack4}\\
&\widetilde\mu_{m,n}^{\star}\left(\left({\sum_{l\in\mathcal{M}} \theta_{l,m} T_{l,n}^{\star} + \eta_{m}}\right)({x_{m,n}^{\star}})^{-1}-1\right)=0,\quad{m\in\mathcal M},\ {n\in\mathcal N},\label{eqn:slack5}\\
&\widetilde\kappa_{1,m,n}^{\star}T_{m,n}^{\star}=0,\quad {m\in\mathcal M},\ {n\in\mathcal N},\label{eqn:k0}\\
&\widetilde\kappa_{2,m,n}^{\star}(T_{m,n}^{\star}-1)=0,\quad {m\in\mathcal M},\ {n\in\mathcal N},\label{eqn:k1}\\
&\widetilde\iota_m^{\star}\left(\sum\limits_{n\in\mathcal N}T_{m,n}^{\star}-K_m\right)=0,\quad {m\in\mathcal M}.\label{eqn:slack8}
\end{align}
\item Zero derivative: 
\begin{align}
&\frac{\partial\mathcal{L}\left(\mathbf{T}^{\star},\mathbf{x}^{\star}, \boldsymbol{\lambda}^{\star}, \boldsymbol{\mu}^{\star}, {\nu_1}^{\star}, {\nu_2}^{\star}, y^{\star}, \widetilde{\boldsymbol\nu}_1^{\star}, \widetilde{\boldsymbol\nu}_2^{\star}, \widetilde\lambda_0^{\star}, \widetilde{\boldsymbol\lambda}^{\star}, \widetilde{\boldsymbol\mu}^{\star}, \widetilde{\boldsymbol\kappa}_1^{\star}, \widetilde{\boldsymbol\kappa}_2^{\star}, \widetilde{\boldsymbol\iota}^{\star}\right)}{\partial y}=\widetilde\lambda_0^{\star}-1=0,\label{eqn:deri-1}\\
&\frac{\partial\mathcal{L}\left(\mathbf{T}^{\star},\mathbf{x}^{\star}, \boldsymbol{\lambda}^{\star}, \boldsymbol{\mu}^{\star}, {\nu_1}^{\star}, {\nu_2}^{\star}, y^{\star}, \widetilde{\boldsymbol\nu}_1^{\star}, \widetilde{\boldsymbol\nu}_2^{\star}, \widetilde\lambda_0^{\star}, \widetilde{\boldsymbol\lambda}^{\star}, \widetilde{\boldsymbol\mu}^{\star}, \widetilde{\boldsymbol\kappa}_1^{\star}, \widetilde{\boldsymbol\kappa}_2^{\star}, \widetilde{\boldsymbol\iota}^{\star}\right)}{\partial\lambda_n}\nonumber\\
&=-\widetilde\nu_{1,n}^{\star}-\widetilde\lambda_0^{\star}\underline a_n+\widetilde\lambda_n^{\star}=0,\quad {n\in\mathcal N},\label{eqn:deri-2}\\
&\frac{\partial\mathcal{L}\left(\mathbf{T}^{\star},\mathbf{x}^{\star}, \boldsymbol{\lambda}^{\star}, \boldsymbol{\mu}^{\star}, {\nu_1}^{\star}, {\nu_2}^{\star}, y^{\star}, \widetilde{\boldsymbol\nu}_1^{\star}, \widetilde{\boldsymbol\nu}_2^{\star}, \widetilde\lambda_0^{\star}, \widetilde{\boldsymbol\lambda}^{\star}, \widetilde{\boldsymbol\mu}^{\star}, \widetilde{\boldsymbol\kappa}_1^{\star}, \widetilde{\boldsymbol\kappa}_2^{\star}, \widetilde{\boldsymbol\iota}^{\star}\right)}{\partial \mu_n}\nonumber\\
&=-\widetilde\nu_{2,n}^{\star}+\widetilde\lambda_0^{\star}\overline a_n-\widetilde\lambda_n^{\star}=0,\quad {n\in\mathcal N},\label{eqn:deri-3}\\
&\frac{\partial\mathcal{L}\left(\mathbf{T}^{\star},\mathbf{x}^{\star}, \boldsymbol{\lambda}^{\star}, \boldsymbol{\mu}^{\star}, {\nu_1}^{\star}, {\nu_2}^{\star}, y^{\star}, \widetilde{\boldsymbol\nu}_1^{\star}, \widetilde{\boldsymbol\nu}_2^{\star}, \widetilde\lambda_0^{\star}, \widetilde{\boldsymbol\lambda}^{\star}, \widetilde{\boldsymbol\mu}^{\star}, \widetilde{\boldsymbol\kappa}_1^{\star}, \widetilde{\boldsymbol\kappa}_2^{\star}, \widetilde{\boldsymbol\iota}^{\star}\right)}{\partial \nu_1}=\widetilde\lambda_0^{\star}-\sum_{n\in\mathcal N}\widetilde\lambda_n^{\star}=0,\label{eqn:deri-4}\\
&\frac{\partial\mathcal{L}\left(\mathbf{T}^{\star},\mathbf{x}^{\star}, \boldsymbol{\lambda}^{\star}, \boldsymbol{\mu}^{\star}, {\nu_1}^{\star}, {\nu_2}^{\star}, y^{\star}, \widetilde{\boldsymbol\nu}_1^{\star}, \widetilde{\boldsymbol\nu}_2^{\star}, \widetilde\lambda_0^{\star}, \widetilde{\boldsymbol\lambda}^{\star}, \widetilde{\boldsymbol\mu}^{\star}, \widetilde{\boldsymbol\kappa}_1^{\star}, \widetilde{\boldsymbol\kappa}_2^{\star}, \widetilde{\boldsymbol\iota}^{\star}\right)}{\partial x_{m,n}}\nonumber\\
&=-\frac{\widetilde\lambda_n^{\star} T_{m,n}^{\star}}{x_{m,n}^{\star}}+\frac{\widetilde\mu_{m,n}^{\star}}{x_{m.n}^{\star}}{\left({\sum_{l\in\mathcal{M}} \theta_{l,m} T_{l,n}^{\star} + \eta_{m}}\right)}=0,\quad {m\in\mathcal M},\ {n\in\mathcal N},\label{eqn:deri-5}\\
&\frac{\partial\mathcal{L}\left(\mathbf{T}^{\star},\mathbf{x}^{\star}, \boldsymbol{\lambda}^{\star}, \boldsymbol{\mu}^{\star}, {\nu_1}^{\star}, {\nu_2}^{\star}, y^{\star}, \widetilde{\boldsymbol\nu}_1^{\star}, \widetilde{\boldsymbol\nu}_2^{\star}, \widetilde\lambda_0^{\star}, \widetilde{\boldsymbol\lambda}^{\star}, \widetilde{\boldsymbol\mu}^{\star}, \widetilde{\boldsymbol\kappa}_1^{\star}, \widetilde{\boldsymbol\kappa}_2^{\star}, \widetilde{\boldsymbol\iota}^{\star}\right)}{\partial T_{m,n}}\nonumber\\
&=-\widetilde\lambda_n^{\star} ({x_{m,n}^{\star}})^{-1}+\sum_{l\in\mathcal M}\widetilde\mu_{l,n}^{\star}\theta_{m,l}({x_{l,n}^{\star}})^{-1}-\widetilde\kappa_{1,m,n}^{\star}+\widetilde\kappa_{2,m,n}^{\star}+\widetilde\iota_m^{\star}=0,\quad {m\in\mathcal M},\ {n\in\mathcal N}.\label{eqn:deri-6}
\end{align}
\end{itemize}

Then, based on the KKT conditions of Problem~\ref{prob:roep}, we show that the primal point $\left(\mathbf{T}^{\star}\right.$, $\mathbf{x}^{\star}$, $\boldsymbol{\lambda}^{\star}$, $\boldsymbol{\mu}^{\star}$, $\left.{\nu_1}^{\star}-{\nu_2}^{\star}\right)$ and the dual point $\left(\widetilde{\boldsymbol\nu}_1^{\star}\right.$, $\widetilde{\boldsymbol\nu}_2^{\star}$, $\widetilde{\boldsymbol\lambda}^{\star}$, $\widetilde{\boldsymbol\kappa}_1^{\star}$, $\widetilde{\boldsymbol\kappa}_2^{\star}$, $\left.\widetilde{\boldsymbol\iota}^{\star}\right)$ satisfy the KKT conditions of Problem~\ref{prob:dualmax}, implying that $\left(\mathbf{T}^{\star}\right.$, $\boldsymbol{\lambda}^{\star}$, $\boldsymbol{\mu}^{\star}$, $\left.{\nu_1}^{\star}-{\nu_2}^{\star}\right)$ is a stationary point of Problem~\ref{prob:dualmax}.
Note that the Lagrangian function of Problem~\ref{prob:dualmax} is given by:
\begin{align}
&\mathcal{L}\left(\mathbf{T},\boldsymbol{\lambda}, \boldsymbol{\mu}, {\nu}, \widetilde{\boldsymbol\nu}_1, \widetilde{\boldsymbol\nu}_2, \widetilde{\boldsymbol\lambda}, \widetilde{\boldsymbol\kappa}_1, \widetilde{\boldsymbol\kappa}_2, \widetilde{\boldsymbol\iota}\right)=-\sum_{n\in\mathcal{N}}{(\lambda_n \underline a_n-\mu_n \overline a_n)}+\nu\nonumber\\
&+\sum_{n\in\mathcal N}\widetilde\nu_{1,n}(-\lambda_n)+\sum_{n\in\mathcal N}\widetilde\nu_{2,n}(-\mu_n)
+\sum_{n\in\mathcal N}\widetilde\lambda_n\left(\lambda_n-\mu_n-\nu-\sum_{m\in\mathcal{M}}{\frac{T_{m,n}}{\left({\sum_{l\in\mathcal{M}} \theta_{l,m} T_{l,n} + \eta_{m}}\right)}}\right)\nonumber\\
&+\sum_{m\in\mathcal M}\sum_{n\in\mathcal N}\widetilde\kappa_{1,m,n}(-T_{m,n})+\sum_{m\in\mathcal M}\sum_{n\in\mathcal N}\widetilde\kappa_{2,m,n}(T_{m,n}-1)+\sum_{m\in\mathcal M}\widetilde\iota_m\left(\sum\limits_{n\in\mathcal N}T_{m,n}-K_m\right). \label{eqn:lag2}
\end{align}
Now, we prove that the primal point $\left(\mathbf{T}^{\star}\right.$, $\mathbf{x}^{\star}$, $\boldsymbol{\lambda}^{\star}$, $\boldsymbol{\mu}^{\star}$, $\left.{\nu_1}^{\star}-{\nu_2}^{\star}\right)$ and the dual point $\left(\widetilde{\boldsymbol\nu}_1^{\star}\right.$, $\widetilde{\boldsymbol\nu}_2^{\star}$, $\widetilde{\boldsymbol\lambda}^{\star}$, $\widetilde{\boldsymbol\kappa}_1^{\star}$, $\widetilde{\boldsymbol\kappa}_2^{\star}$, $\left.\widetilde{\boldsymbol\iota}^{\star}\right)$ satisfy the following KKT conditions.
\begin{itemize}
\item Complementary slackness: It is clear that $\left(\mathbf{T}^{\star}\right.$, $\mathbf{x}^{\star}$, $\boldsymbol{\lambda}^{\star}$, $\boldsymbol{\mu}^{\star}$, $\left.{\nu_1}^{\star}-{\nu_2}^{\star}\right)$ and $\left(\widetilde{\boldsymbol\nu}_1^{\star}\right.$, $\widetilde{\boldsymbol\nu}_2^{\star}$, $\widetilde{\boldsymbol\lambda}^{\star}$, $\widetilde{\boldsymbol\kappa}_1^{\star}$, $\widetilde{\boldsymbol\kappa}_2^{\star}$, $\left.\widetilde{\boldsymbol\iota}^{\star}\right)$ satisfy \eqref{eqn:lambdacs}, \eqref{eqn:mucs}, \eqref{eqn:k0} and \eqref{eqn:k1}.
\item Zero derivative: Substituting $\widetilde\lambda_0^{\star}=1$ into \eqref{eqn:deri-2}, \eqref{eqn:deri-3} and \eqref{eqn:deri-4}, we have:
\begin{align}
&\frac{\partial\mathcal{L}\left(\mathbf{T}^{\star},\boldsymbol{\lambda}^{\star}, \boldsymbol{\mu}^{\star}, {\nu}^{\star}, \widetilde{\boldsymbol\nu}_1^{\star}, \widetilde{\boldsymbol\nu}_2^{\star}, \widetilde{\boldsymbol\lambda}^{\star}, \widetilde{\boldsymbol\kappa}_1^{\star}, \widetilde{\boldsymbol\kappa}_2^{\star}, \widetilde{\boldsymbol\iota}^{\star}\right)}{\partial\lambda_n}=-\widetilde\nu_{1,n}^{\star}-\underline a_n+\widetilde\lambda_n^{\star}=0,\quad {n\in\mathcal N},\label{eqn:deri-10}\\
&\frac{\partial\mathcal{L}\left(\mathbf{T}^{\star},\boldsymbol{\lambda}^{\star}, \boldsymbol{\mu}^{\star}, {\nu}^{\star}, \widetilde{\boldsymbol\nu}_1^{\star}, \widetilde{\boldsymbol\nu}_2^{\star}, \widetilde{\boldsymbol\lambda}^{\star}, \widetilde{\boldsymbol\kappa}_1^{\star}, \widetilde{\boldsymbol\kappa}_2^{\star}, \widetilde{\boldsymbol\iota}^{\star}\right)}{\partial \mu_n}=-\widetilde\nu_{2,n}^{\star}+\overline a_n-\widetilde\lambda_n^{\star}=0,\quad {n\in\mathcal N}.\label{eqn:deri-20}\\
&\frac{\partial\mathcal{L}\left(\mathbf{T}^{\star},\boldsymbol{\lambda}^{\star}, \boldsymbol{\mu}^{\star}, {\nu}^{\star}, \widetilde{\boldsymbol\nu}_1^{\star}, \widetilde{\boldsymbol\nu}_2^{\star}, \widetilde{\boldsymbol\lambda}^{\star}, \widetilde{\boldsymbol\kappa}_1^{\star}, \widetilde{\boldsymbol\kappa}_2^{\star}, \widetilde{\boldsymbol\iota}^{\star}\right)}{\partial \nu}=1-\sum_{n\in\mathcal N}\widetilde\lambda_n^{\star}=0. \label{eqn:deri-30}
\end{align}
Substitute $\widetilde\lambda_0^{\star}=1$ into \eqref{eqn:deri-2}, we have $\widetilde\lambda_n^{\star}=\underline a_n+\widetilde\nu_{1,n}^{\star}>0$. By $\widetilde\lambda_n^{\star}>0$ and \eqref{eqn:deri-5}, we have 
\begin{align}
\widetilde\mu_{m,n}^{\star}=\frac{\widetilde\lambda_n T_{m,n}^{\star}}{\left({\sum_{l\in\mathcal{M}} \theta_{l,m} T_{l,n}^{\star} + \eta_{m}}\right)}>0.\label{eqn:mueq}
\end{align}
In addtion, by $\widetilde\mu_{m,n}^{\star}>0$ and \eqref{eqn:slack5}, we have 
\begin{align}
{x^{\star}_{m,n}}=\left({\sum_{l\in\mathcal{M}} \theta_{l,m} T_{l,n}^{\star} + \eta_{m}}\right).\label{eqn:xeq}
\end{align}
Substituting \eqref{eqn:mueq} and \eqref{eqn:xeq} into \eqref{eqn:deri-6}, we have:
\begin{align}
&\frac{\partial\mathcal{L}\left(\mathbf{T}^{\star},\boldsymbol{\lambda}^{\star}, \boldsymbol{\mu}^{\star}, {\nu}^{\star}, \widetilde{\boldsymbol\nu}_1^{\star}, \widetilde{\boldsymbol\nu}_2^{\star}, \widetilde{\boldsymbol\lambda}^{\star}, \widetilde{\boldsymbol\kappa}_1^{\star}, \widetilde{\boldsymbol\kappa}_2^{\star}, \widetilde{\boldsymbol\iota}^{\star}\right)}{\partial T_{m,n}}\nonumber\\
&=-\widetilde\lambda_n^{\star}\nabla_{T_{m,n}}\frac{T_{m,n}^{\star}}{\left({\sum_{l\in\mathcal{M}} \theta_{l,m} T_{l,n}^{\star} + \eta_{m}}\right)}-\widetilde\kappa_{1,m,n}^{\star}+\widetilde\kappa_{2,m,n}^{\star}+\widetilde\iota_m^{\star}=0,\ {m\in\mathcal M},\ {n\in\mathcal N}.\label{eqn:deri-40}
\end{align}
Note that \eqref{eqn:deri-10}, \eqref{eqn:deri-20}, \eqref{eqn:deri-30} and \eqref{eqn:deri-40} are the zero derivative conditions of Problem~\ref{prob:dualmax}.
\item Feasibility: Given \eqref{eqn:dual20}, \eqref{eqn:dual30}, \eqref{eqn:provprob4}, \eqref{eqn:provprob5} and \eqref{eqn:dual-feasible}, to show that $\left(\mathbf{T}^{\star}\right.$, $\mathbf{x}^{\star}$, $\boldsymbol{\lambda}^{\star}$, $\boldsymbol{\mu}^{\star}$, $\left.{\nu_1}^{\star}-{\nu_2}^{\star}\right)$ and $\left(\widetilde{\boldsymbol\nu}_1^{\star}\right.$, $\widetilde{\boldsymbol\nu}_2^{\star}$, $\widetilde{\boldsymbol\lambda}^{\star}$, $\widetilde{\boldsymbol\kappa}_1^{\star}$, $\widetilde{\boldsymbol\kappa}_2^{\star}$, $\left.\widetilde{\boldsymbol\iota}^{\star}\right)$ satisfy the feasibility conditions of Problem~\ref{prob:dualmax}, it remains to show:
\begin{align}
&\sum_{m\in\mathcal{M}}{\frac{T_{m,n}^\star}{\sum_{l\in\mathcal{M}}{\theta_{l,m}T_{l,n}^\star}+\eta_{m}}}+\mu_n^\star-\lambda_n^\star+\nu^\star_1-\nu^\star_2=0,\quad n\in\mathcal{N},\label{eqn:dual100}\\
&\sum\limits_{n\in\mathcal N}T_{m,n}^\star=K_m,\quad m\in\mathcal M. \label{eqn:T200}
\end{align}
Note that \eqref{eqn:dual100} can be easily shown by $\widetilde\lambda_n^{\star}>0$, \eqref{eqn:slack4} and \eqref{eqn:xeq}. Now, we show \eqref{eqn:T200} as below.
If $\widetilde\kappa_{2,m,n}^{\star}\neq0$ for all $n\in\mathcal N$, by \eqref{eqn:k1}, we have $T^{\star}_{m,n}=1$ for all $n\in\mathcal N$, implying $\sum_{n\in\mathcal N}T^{\star}_{m,n}=N$, which is contradictory with $\sum_{n\in\mathcal N}T^{\star}_{m,n}\leq K_m$. Thus, there exists $n\in\mathcal N$, such that $\widetilde\kappa_{2,m,n}^{\star}=0$. Substituting $\widetilde\kappa_{2,m,n}^{\star}=0$ into $\eqref{eqn:deri-40}$, we have: 
\begin{align}
\widetilde\iota_m^{\star}=\widetilde\lambda_n^{\star}\nabla_{T_{m,n}}\frac{T_{m,n}^{\star}}{\left({\sum_{l\in\mathcal{M}} \theta_{l,m} T_{l,n}^{\star} + \eta_{m}}\right)}+\widetilde\kappa_{1,m,n}^{\star}>0. \nonumber
\end{align}
By $\widetilde\iota_m^{\star}>0$ and \eqref{eqn:slack8}, we have \eqref{eqn:T200}.
\end{itemize}
Therefore, $\left(\mathbf{T}^{\star}\right.$, $\mathbf{x}^{\star}$, $\boldsymbol{\lambda}^{\star}$, $\boldsymbol{\mu}^{\star}$, $\left.{\nu_1}^{\star}-{\nu_2}^{\star}\right)$ and $\left(\widetilde{\boldsymbol\nu}_1^{\star}\right.$, $\widetilde{\boldsymbol\nu}_2^{\star}$, $\widetilde{\boldsymbol\lambda}^{\star}$, $\widetilde{\boldsymbol\kappa}_1^{\star}$, $\widetilde{\boldsymbol\kappa}_2^{\star}$, $\left.\widetilde{\boldsymbol\iota}^{\star}\right)$ satisfy the KKT conditions of Problem~\ref{prob:dualmax}.

\section*{Appendix D: Proof of Theorem~\ref{thm:conv-sto}}
We show that the assumptions in~\cite[Theorem~1]{7412752} are satisfied. 
\begin{itemize}
\item The constraint set of Problem~\ref{prob:sto} determined by \eqref{eqn:T1} and \eqref{eqn:T2} is compact and convex. Thus, Assumption a) in~\cite[Theorem~1]{7412752} is satisfied.
\item It is clear that $q_m(\boldsymbol\xi,\mathbf T)$ is smooth on the constraint set of Problem~\ref{prob:sto} for any given $\boldsymbol\xi$, and hence it is continuously differentiable and its derivative is Lipschitz continuous. Thus, Assumption b) in~\cite[Theorem~1]{7412752} is satisfied.
\item Random variables $\boldsymbol\xi^0,\boldsymbol\xi^1,\dots$ are bounded and identically distributed. Thus, Assumption c) in~\cite[Theorem~1]{7412752} is satisfied.
\end{itemize}
Therefore, Theorem~\ref{thm:conv-sto} readily follows from~\cite[Theorem~1]{7412752}.

\end{document}